\documentclass[fleqn,usenatbib]{mnras}

\usepackage{newtxtext,newtxmath}
\usepackage{mathrsfs}
\usepackage{multirow}

\usepackage[T1]{fontenc}
\usepackage{ae,aecompl}

\usepackage{graphicx}	
\usepackage{amsmath}	
\usepackage[x11names]{xcolor}

\title[Constraining $H_{0}$ from the proper motion]{Constraining the Hubble constant and its lower limit from the proper motion of extragalactic radio jets}

\author[Hsiao et al. 2022]{Tiger Yu-Yang Hsiao,$^{1,2}$
Tomotsugu Goto,$^{2}$
Tetsuya Hashimoto,$^{3}$
Daryl Joe D. Santos,$^{4}$
Yi Hang Valerie Wong,$^{2,5}$
\newauthor
Seong Jin Kim,$^{2}$
Bjorn Jasper R. Raquel,$^{3,6}$
Simon C.-C. Ho,$^{2}$
Bo-Han Chen,$^{2,7}$
Ece Kilerci,$^{8}$
\newauthor
Ting-Yi Lu,$^{9,10}$
Alvina Y. L. On,$^{2,11}$
Yu-Wei Lin,$^{2,7}$
and Cossas K.-W. Wu$^{2,7}$
\\
$^{1}$Department of Physics and Astronomy, Johns Hopkins University, Baltimore, MD 21218, USA\\
$^{2}$Institute of Astronomy, National Tsing Hua University, 101, Section 2. Kuang-Fu Road, Hsinchu, 30013, Taiwan (R.O.C.)\\
$^{3}$Department of Physics, National Chung Hsing University, No. 145, Xingda Rd., South Dist., Taichung, 40227, Taiwan (R.O.C.)\\
$^{4}$Max Planck Institute for Extraterrestrial Physics, Gießenbachstraße 1, 85748 Garching, Germany  \\
$^{5}$Department of Astrophysical and Planetary Science, University of Colorado Boulder, CO 80309, USA\\
$^{6}$Department of Earth and Space Sciences, Rizal Technological University, Boni Avenue, Mandaluyong, 1550 Metro Manila, Philippines\\
$^{7}$Department of Physics, National Tsing Hua University, 101, Section 2. Kuang-Fu Road, Hsinchu, 30013, Taiwan (R.O.C.)\\
$^{8}$Sabanc{\i} University, Faculty of Engineering and Natural Sciences, 34956, Istanbul, Turkey\\
$^{9}$Niels Bohr Institute, University of Copenhagen, Jagtvej 128, København N, DK-2200, Denmark\\
$^{10}$Cosmic Dawn Center (DAWN)\\
$^{11}$Mullard Space Science Laboratory, University College London, Holmbury St Mary, Surrey RH5 6NT, UK}

\date{Accepted 2022 September 6. Received 2022 September 2; in original form 2022 March 3}

\pubyear{2022}

\begin{document}
\label{firstpage}
\pagerange{\pageref{firstpage}--\pageref{lastpage}}
\maketitle
\begin{abstract}
The Hubble constant ($H_{0}$) is a measurement to describe the expansion rate of the Universe in the current era.
However, there is a $4.4\sigma$ discrepancy between the measurements from the early Universe and the late Universe.
In this research, we propose a model-free and distance-free method to constrain $H_{0}$.
Combining Friedman-Lemaître-Robertson-Walker cosmology with geometrical relation of the proper motion of extragalactic jets, the lower limit ($H_{\rm 0,min}$) of $H_{0}$ can be determined using only three cosmology-free observables: the redshifts of the host galaxies, as well as the approaching and receding angular velocities of radio jets.
Using these, we propose to use the Kolmogorov-Smirnov test (K-S test) between cumulative distribution functions of $H_{\rm 0,min}$ to differentiate cosmology.
We simulate 100, 200, and 500 extragalactic jets with 3 levels of accuracy of the proper motion ($\mu_{a}$ and $\mu_{r}$), at 10\%, 5\%, and 1\%, corresponding to the accuracies of the current and future radio interferometers.
We perform K-S tests between the simulated samples as theoretical distributions with different $H_{0}$ and {power-law index} of velocity distribution of jets and mock observational data. 
Our result suggests increasing sample sizes leads to tighter constraints on both {power-law index} and the Hubble constant at moderate accuracy (i.e., $10\%$ and $5\%$) while at $1\%$ accuracy, increasing sample sizes leads to tighter constraints on {power-law index} more. Improving accuracy results in better constraints in the Hubble constant compared with the {power-law index} in all cases but it alleviates the degeneracy.

\end{abstract}


\maketitle


\section{Introduction}
\label{introduction}
{Approximately a century ago, the Universe was found to be expanding \citep{Hubble1929}.}
The Hubble parameter, $H(z)$, describes the expansion rate of the Universe.
Its value in the current era ($z=0$), known as the Hubble constant, is denoted by $H_{0}$.
Numerous measurements of the Hubble constant {observed} from different methods have been proposed to measure the values during the past few decades.
For instance, Cosmic Microwave Background (CMB), remnant of the Big Bang in the early Universe, which is also a standard ruler as a cosmological probe.
One of the latest measurements from \citet{Planck2020} suggested $H_{0}=67.27\pm0.60\,{\rm km\,s^{-1}\,Mpc^{-1}}$.
Additionally, the method of the distance ladder, such as Cepheid variables in the late Universe, can be used to constrain the Hubble constant directly.
The Supernova, H0, for the Equation of State \citep[SH0ES;][]{Riess2021}, which {adopted the method of distance ladder. They utilise Cepheids to calibrate 42 Type Ia supernovae in the same host galaxies, obtained a value of $H_{0}=73.04\pm1.04\,{\rm km\,s^{-1}\,Mpc^{-1}}$.}
Other methods such as using baryon acoustic oscillations \citep[e.g.,][]{Eisenstein2005,Cuceu2019}, big bang nucleosynthesis \citep[e.g.,][]{Cuceu2019,Seto2021}, strong gravitational lensing of quasars \citep[e.g.,][]{Wong2020}, water masers \citep[e.g.,][]{Herrnstein1999,Humphreys2013}, gravitationally lensed supernova \citep[e.g.,][]{Refsdal1964,Ferrero2018} were also utilised to measure the Hubble constant.

However, as we obtain more precise measurements, the results conducted from the early Universe further reveal an inconsistency with the Hubble constant from the late Universe.
This discrepancy is known as "Hubble tension" and indicates a tension beyond $4.4\sigma$ \citep[e.g.,][]{Verde2019}.
The reason for the tension is still under debate.
One possible theory is that, there may be hints of new physics beyond the $\Lambda$CDM model.
For instance, there are popular theories such as the early dark energy \citep[e.g.,][]{Poulin2019}, acoustic dark energy \citep[e.g.,][]{Yin2020}, or importing sterile neutrino \citep[e.g.,][]{Adhikari2017} trying to tackle this issue.
In terms of observation, aside from observing standard candles and standard rulers, several new methods have also been proposed to address this discrepancy for the next generation.
For example, 
since gravitational waves (GWs) act as standard sirens, an independent way to infer luminosity distance from their amplitudes \citep{Schutz1986}, some studies suggested constraining cosmological parameters using GWs in the future \citep[e.g.,][]{Abbott2017,Chen2018}.
Besides, fast radio bursts (FRBs), radio pulses with millisecond time scale, can also be used to constrain cosmology thanks to its unique observable quantities, dispersion measure \citep[e.g.][]{Li2018,Wu2021} and duration-luminosity relation \citep[e.g.,][Hsiao et al. in prep]{Hashimoto2019}.

Relativistic jets are one of the energetic phenomena which produce piercing matter from regions near compact objects such as black holes and neutron stars.
These robust plasma flows are {expected to be} along the axis of rotation of the host object.
Although the relation between the
accretion disc and the relativistic jets is still in debate, it is believed
that the jets are driven by the tangled magnetic field \citep[e.g.,][]{Blandford1977,Hawley2002,McKinney2004}.
Previous studies revealed that it is possible to extract the information of the Hubble constant from the proper motion of jets \citep[e.g.,][]{Lynden1977,Taylor1997,Qin1999,Lu2021} and with GWs \citep[e.g.,][]{Hotokezaka2019}.
Therefore, the observation of jets have potential to help us relieve the Hubble tension.

In this paper, we adopt the method proposed from \citet{Qin1999} \citep[see also][]{Taylor1997} to calculate the lower limit of the Hubble constant through the Friedman-Lema\^itre-Robertson-Walker (FLRW) cosmology and geometrical proper motion of extragalactic radio jets.
With only the redshifts, and the receding and approaching angular velocities, the lower limit of the Hubble constant can be determined.
The first measurement originated from \citet{Taylor1997}, suggesting $H_{\rm 0,min}=37\,{\rm km\,s^{-1}\,Mpc^{-1}}$.
\citet{Qin1999} adopted a similar method and suggested $27.08\,{\rm km\,s^{-1}\,Mpc^{-1}}<H_{\rm 0,min}<53.15\,{\rm km\,s^{-1}\,Mpc^{-1}}$.
\citet{Lu2021} made use of an extragalactic radio source, NGC 1052, located at $z=0.005$.
Their results suggested the lower limit of the Hubble constant is $H_{\rm 0,min}=51.5\pm2.3\,{\rm km\,s^{-1}\,Mpc^{-1}}$, {which successfully follows the method of \citet{Taylor1997} in constraining the Hubble constant.}
Apparently, the latest constraint from proper motion is not enough to relieve the Hubble tension.
Therefore, in this paper, {we investigate how well $H_{0}$ can be constrained from $H_{\rm 0,min}$ distributions by simulating samples of observed speeds in nearby bi-directional AGN jets.} 
Furthermore, we perform Kolmogorov-Smirnov tests \citep[K-S test;][]{KS} to compare the cumulative distribution function (CDF) of the mock data to theoretical distribution.
K-S test aims to compare if two distributions are drawn from the same distributions, in order to obtain further information of $H_{0}$ and {power-law index} $k$ based on the cumulative distribution function of $H_{\rm 0,min}$.
 
The structure of this paper is as follows.
We describe the theoretical framework and our method in Sec. \ref{Method}.
In Sec. \ref{Discussion}, we present the results and, distributions of $H_{\rm 0,min}$ from our simulation according to two different cosmology, and further constrain $H_{0}$ and $k$ from K-S tests.
The conclusion of this study is summarised in Sec. \ref{Conclusion}.
We assume concordance cosmology with ($\Omega_m$, $\Omega_{\Lambda}$, $h$) = ($0.3$, $0.7$, $0.7$), unless otherwise mentioned.

\section{Methodology}
\label{Method}
According to FLRW cosmology, the Hubble law for a nearby extragalactic source with $z\ll1$ can be written as follows:
\begin{equation}
\label{FLRW}
    \frac{D_{L}}{(1+z)}\approx\frac{cz}{H_{0}},
\end{equation}
where $D_{L}$ is the luminosity distance and $c$ is the speed of light. 

\begin{figure}
    	\includegraphics[width=\columnwidth]{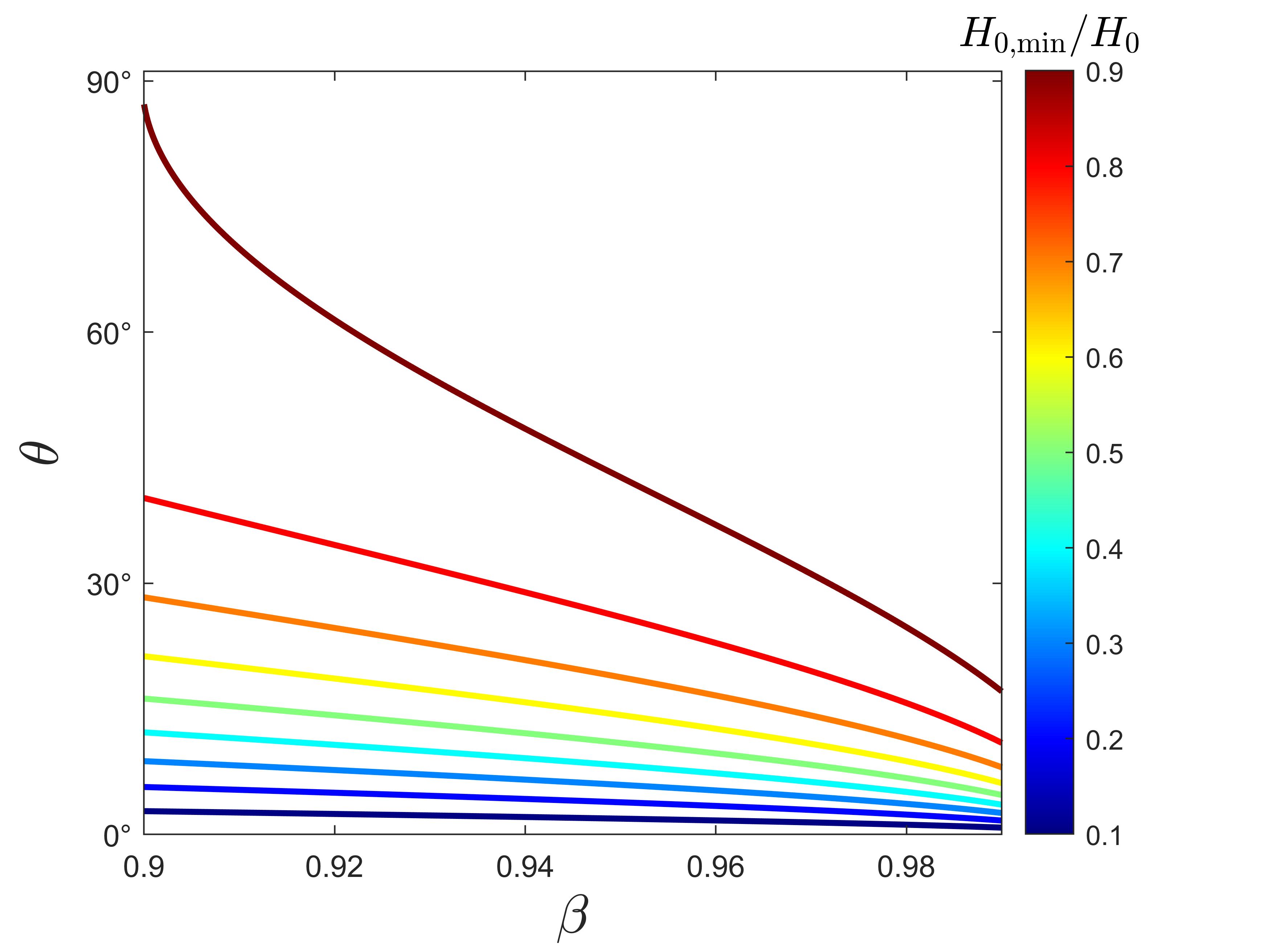}
    	\caption{{Contours of $H_{\rm 0,min}$/$H_{0}$ as a function of $\theta$ and $\beta$ with fixed $z=0.001$ and $H_{0}=70\,{\rm km\,s^{-1}\,Mpc^{-1}}$.}}
    	
        \label{thetabeta}
        
    \end{figure}

In terms of the geometry, considering a bi-symmetric  relativistic jet, the proper motion of the receding and approaching jets can be illustrated as \citep[e.g.,][]{Rees1966,Behr1976,Blandford1977b,Blandford1979,Mirabel1994}:
\begin{equation}
\label{pro}
    \mu_{r,a}=\frac{\beta {\rm sin}\theta}{1\pm\beta {\rm cos}\theta}\frac{c(1+z)}{D_{L}},
\end{equation}
where $\mu_{r}$ and $\mu_{a}$ are the proper motions of receding and approaching jets, respectively.
$\beta$ is the ratio between the jet velocity and the speed of light ($v/c$) while $\theta$ is the angle between {the velocity vector of the approaching jet} and the line of sight.
Eqs. \ref{FLRW} and \ref{pro} yield \citep[see also][]{Qin1999,Lu2021}:
\begin{equation}
    H_{0}\simeq\frac{2\mu_{\rm a}\mu_{\rm r}z}{\sqrt{\beta^{2}(\mu_{\rm a}+\mu_{\rm r})^{2}-(\mu_{\rm a}-\mu_{\rm r})^{2}}}.
\end{equation}
Lastly, since velocity must be smaller than the speed of light ($\beta<1$), we obtain a relation among the lower limit of the Hubble constant, proper motion, and redshift:
\begin{equation}
\label{H0min}
H_{\rm 0,min}=z\sqrt{\mu_{a}\mu_{r}}.
\end{equation}
{We note that there are assumptions that the jets are straight and oppositely directed, and have identical bulk flow speed.
If the velocity is higher (especially close to the speed of light), the $H_{\rm 0,min}$ will be close to the true $H_{0}$ at any fixed $\theta$.
Contours of how $\theta$ and $\beta$ will change the fraction of $H_{\rm 0,min}$/$H_{0}$ is shown in Fig. \ref{thetabeta}.}
In principle, if there is a fraction of the speed of jets that is close to the speed of light, we expect to obtain a $H_{\rm 0,min}$ which is close to $H_{0}$ from those samples.
Therefore, this method provides a model-free and distance-free way to constrain the Hubble constant only from geometry and FLWR cosmology, and may be able to alleviate $H_{0}$ tension.

        \begin{figure}
    	\includegraphics[width=\columnwidth]{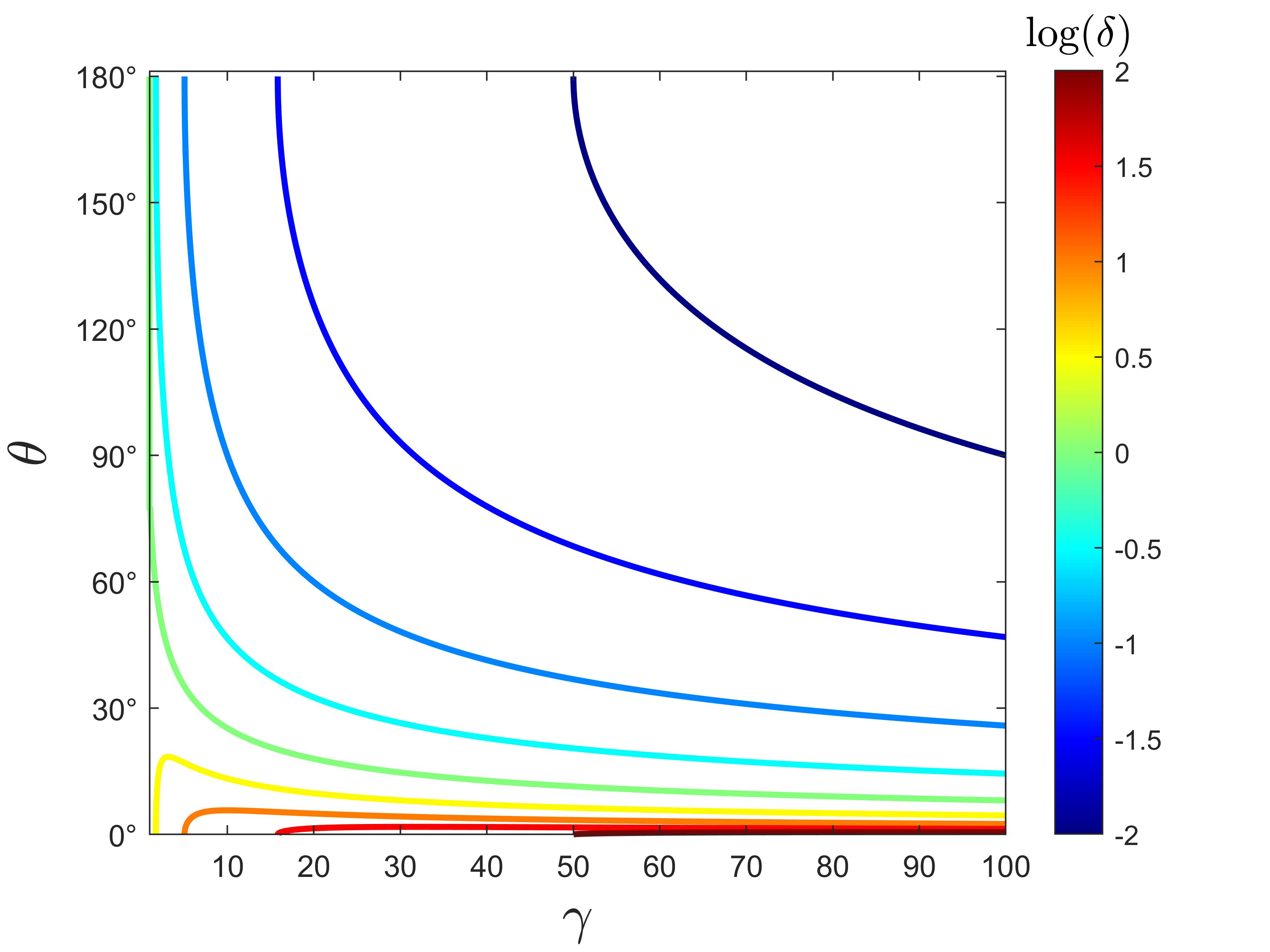}
    	\caption{{Contours of the Doppler factor as a function of $\theta$ and $\gamma$.}}
    	
        \label{dopplerfactor}
        
    \end{figure}    


{We note that due to the relativistic beaming effect (or the Doppler boosting), the receding jets are not easily observed \citep[e.g.,][]{Wilkinson1977,Bridle1984,Sparks1992,Laing2002}.
For an ideal relativistic jet, the observed luminosity can be expressed as \citep[e.g.,][]{Blandford1979,Cohen2007}
\begin{equation}
\label{dopplerbeaming}
L_{\rm obs}=L_{\rm int}\delta^{(p-\alpha)},
\end{equation}
where $L_{\rm obs}$ is the observed luminosity and $L_{\rm int}$ is the intrinsic luminosity.
$\delta$ is the Doppler factor ($\delta=\gamma^{-1}(1-\beta {\rm cos}(\theta))^{-1}$), and $\gamma$ is the Lorentz factor ($\gamma=1/\sqrt{1-\beta^{2}}$).
$\alpha$ is the spectral index.
$p$ is the Doppler boost exponent \citep[for more detail, see][]{Cohen2007}.
Contours of Dopper factor $\delta$ as a function of $\gamma$ and $\theta$ are shown in Fig. \ref{dopplerfactor}.
For example, in the catalogue of Monitoring Of Jets in Active galactic nuclei with VLBA Experiments (MOJAVE) program \citep{Lister2005,Homan2021}, the two-sided jets account for $\sim18\%$ of total sources.
In this research, we simply assume the samples are those which have two-sided jets. 
}

\begin{figure}
    	\includegraphics[width=\columnwidth]{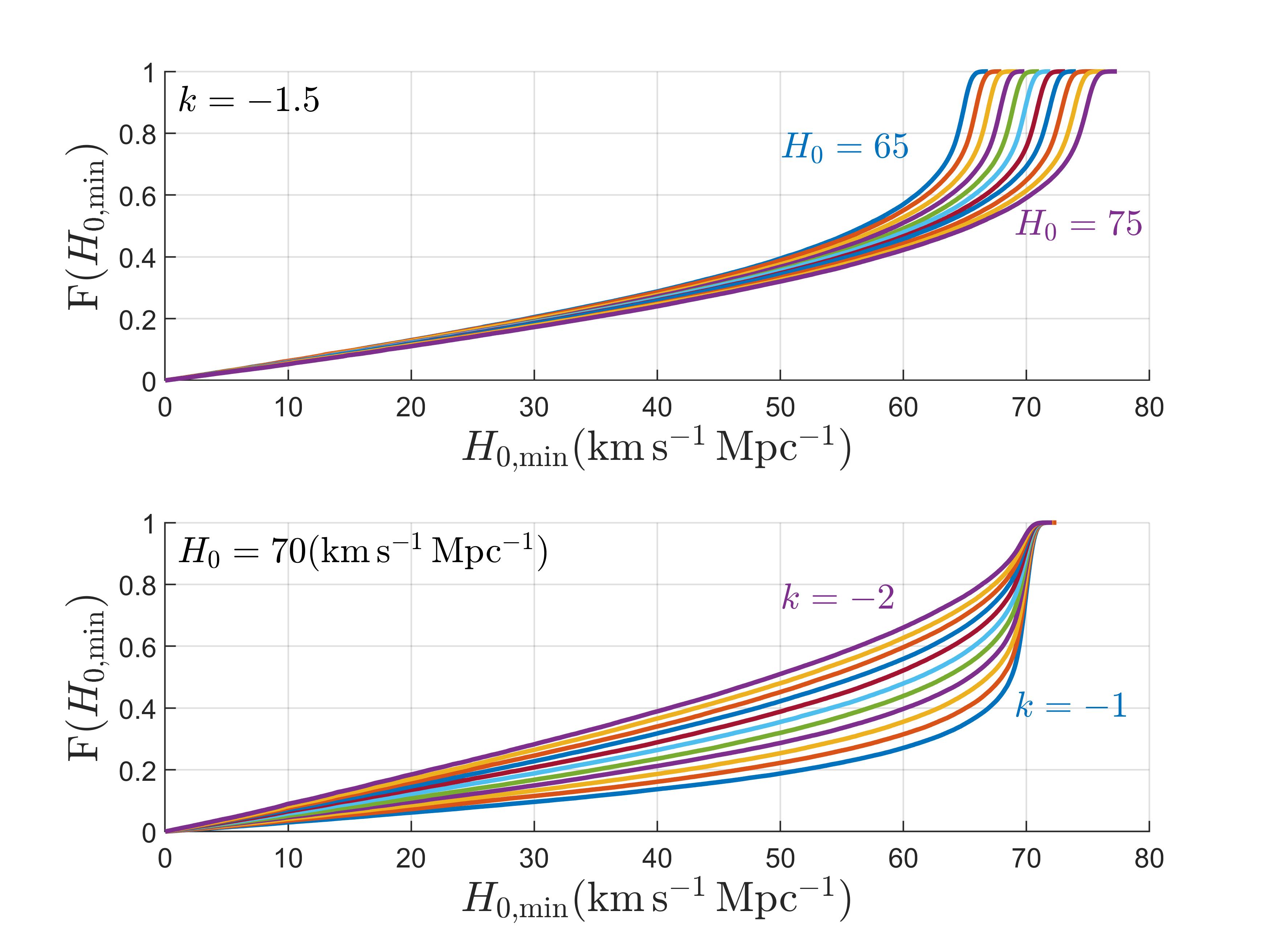}
    	\caption{{CDFs of $H_{\rm 0,min}$ under different prior $H_{0}$ (the upper panel) and $k$ (the lower panel).}}
    	
        \label{theo}
        
    \end{figure}

We propose a new method {to test how well we can recover $H_{0}$ and the power-law index, $k$ (see the explanation in the next paragraph) from the distribution of $H_{\rm 0,min}$ gave a fixed sample size and a typical measurement error in the future.
In this process, in addition to the high Lorentz factor jets, we also utilise lower Lorentz factor jets, in fact, we use the whole distribution to estimate $H_{0}$. In this way, the usable sample size becomes much larger.}
By conducting the K-S test between {mock} observational data {(fixed $H_{0}=70\,{\rm km\,s^{-1}\,Mpc^{-1}}$ and $k=-1.5$) and the theoretical CDFs (with different prior $H_{0}$ and prior $k$), the p-value, a similarity of two profiles, can be represented as probabilities of $H_{0}$ and $k$.}
{We simulate a large number of jets, i.e., $10^{5}$, with $H_{\rm 0,min}$, as theoretical distributions for each point between $65\,{\rm km\,s^{-1}\,Mpc^{-1}}<H_{0}<75\,{\rm km\,s^{-1}\,Mpc^{-1}}$ and $-1<k<-2$.
Fig. \ref{theo} shows how the true $H_{0}$ and $k$ changes the features of CDFs.
}
{Figs. \ref{methodH0} and \ref{methodk} show how our idea works.}
{1\% error on the measurement of proper motions and 100 sources are assumed for the mock observational data here.}
Since {the feature of} a CDF of $H_{\rm 0,min}$ changes when $H_{0}$ changes, p-values from K-S tests will be small when $H_{0}$ of the observational data does not match that of the theoretical one, and vice versa.
{For example, in the upper panels of Fig. \ref{methodH0}, we compare the similarity via the K-S test of the mock observational data (shown in blue curves) and the theoretical curves (the orange, the yellow, and the green curves are assumed $H_{0}=65$, $H_{0}=70$, and $H_{0}=75\,{\rm km\,s^{-1}\,Mpc^{-1}}$, respectively).
The similarity between mock observational data and the yellow curve is the highest (p-value is the highest).
Therefore, we infer the p-value as a probability as a function of $H_{0}$ to constrain $H_{0}$.}
After we conduct K-S tests between the mock observational data and the theoretical CDFs under different $H_{0}$ and $k$, we are able to constrain $H_{0}$ from the statistical p-value as a function of $H_{0}$ and $k$.

{For both mock observational data and the theoretical data,} we consider a distribution of the velocity following the power-law distribution \citep[e.g.,][]{Lister1997,Lister2003,Cara2008,Ajello2012,Lister2016,Yuan2018}:
\begin{equation}
    P_{\gamma}(\gamma)=C\gamma^{k},
\end{equation}
where $C$ is a normalised factor, $k$ is a {power-law index}.
We {assume $k=-1.5$ for the mock observational data}, which is the best fit suggested by superluminal motions \citep{Lister1997}.
The interval of the velocity is set to be $1.01\leq\gamma\leq100$ \citep[e.g.,][]{Yuan2018}.
{The samples are randomly distributed within the spatial volume (i.e., population has constant space density) between $0<z<0.02$ and $0<\theta<\pi/2$.
Note that for 3-dimensionally oriented two-sided jets, random jet viewing angles in the population $P(\theta,\theta+d\theta) \propto \sin(\theta)d\theta$ \citep[e.g.,][]{Law2009}.
The effects of Doppler orientation bias may be minimal for $z<0.02$ and the effects are ignored in the simulation for simplification.}
{The $H_{0}$ of mock observational data is set to be $70\,{\rm km\,s^{-1}\,Mpc^{-1}}$.}
As a result, we calculate $\mu_{r}$ and $\mu_{a}$ for each jet according to Eq. \ref{pro}.
After the simulation, we construct $H_{\rm 0,min}$ based on Eq. \ref{H0min}.
As for the uncertainty, it depends on {not only} the angular resolution and the distance to the object {but also depends strongly on total time coverage and cadence, the brightness temperature of the moving feature, the relative stability of the core (reference point), confusion with nearby jet features, and possible accelerations/non-linear motions \citep[e.g.,][]{Lister2021,Weaver2022}.}
{For simplicity,} we simulate $\mu_{a}$ and $\mu_{r}$ with the observational uncertainty of 10\%, 5\%, and 1\%, {which is based on the uncertainties in the MOJAVE program \citep{Homan2021}.}
In addition, we assume three sets of jets: 100, 200, and 500 jets, which correspond to $\sim20\%$, $\sim40\%$ and $\sim100\%$ of the number of the jets in {the MOJAVE program. }
Furthermore, there are $\sim120$ radio galaxies at $z<0.02$ with jets/lobes structure from 2 Micron All-Sky Survey (2MASS) Redshift Survey \citep{Velzen2012}.
We expect that observing the nearby sources with higher sensitivity radio telescopes will increase the sample size.
To estimate errors of $H_{\rm 0,min}$, we conducted each simulation 1,000 times by adding randomized errors (i.e., 10\%, 5\%, and 1\%) to $\mu_{a}$ and $\mu_{r}$.


\section{Results and Discussion}
\label{Discussion}

     \begin{figure*}
    	\includegraphics[width=\textwidth]{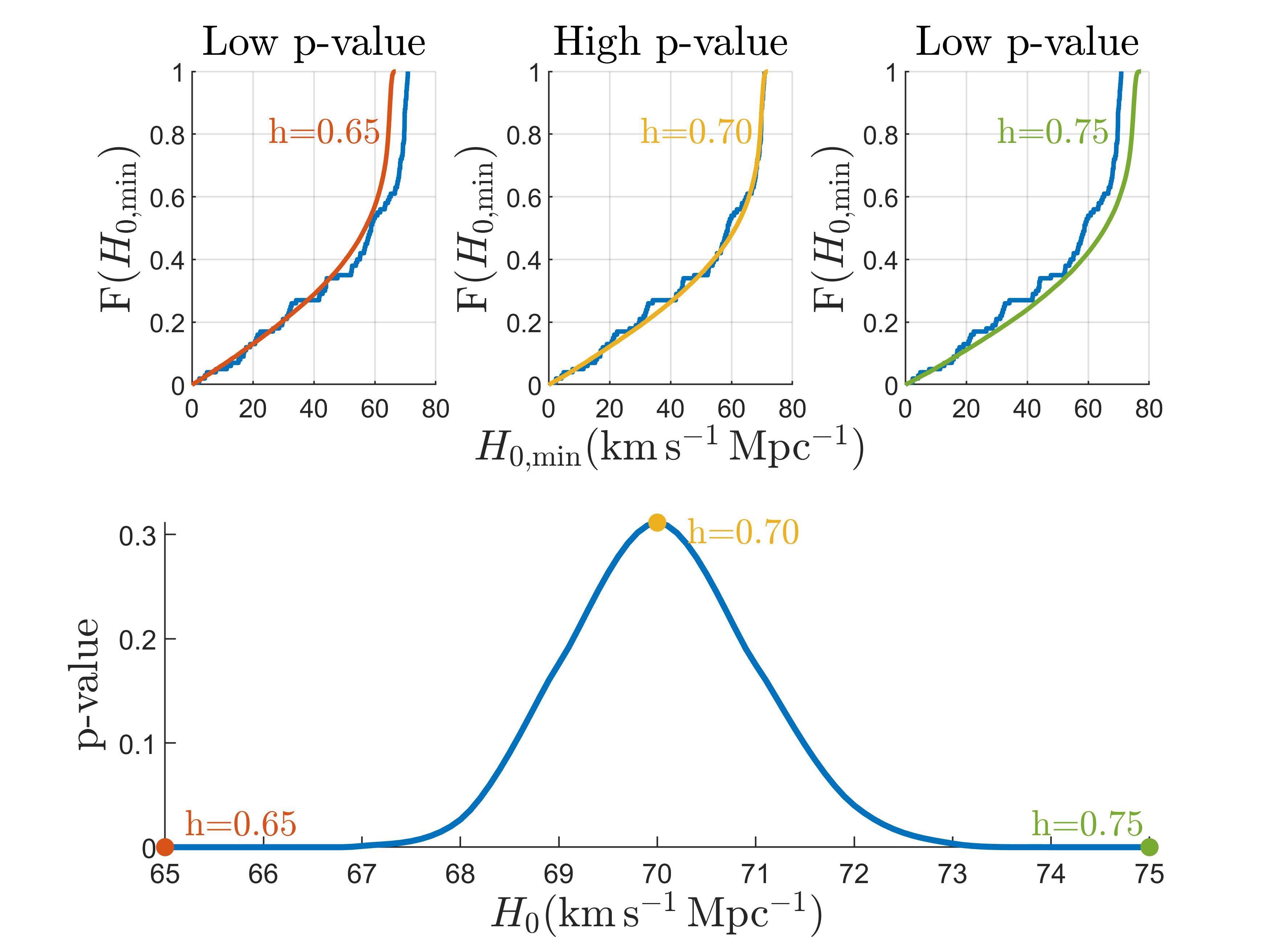}
    	\caption{{This figure shows how our method works. The upper three panels show CDFs of the theoretical profile with different assumed $H_{0}$, $65$ (the orange curve), $70$ (the yellow curve), and $75\,{\rm km\,s^{-1}\,Mpc^{-1}}$ (the green curve) and the mock observational data (the blue curve; assumed $H_{0}=70\,{\rm km\,s^{-1}\,Mpc^{-1}}$). {1\% error on the measurement of proper motions and 100 sources are assumed for the mock observational data here.} The lower panel indicates the p-value of the K-S tests between mock observational data and theoretical data as a function of assumed $H_{0}$. The orange dot, the yellow dot, and the green dot represent the p-value of the K-S tests in the three upper panels.}}
    	
        \label{methodH0}
        
    \end{figure*}   
    
     \begin{figure*}
    	\includegraphics[width=\textwidth]{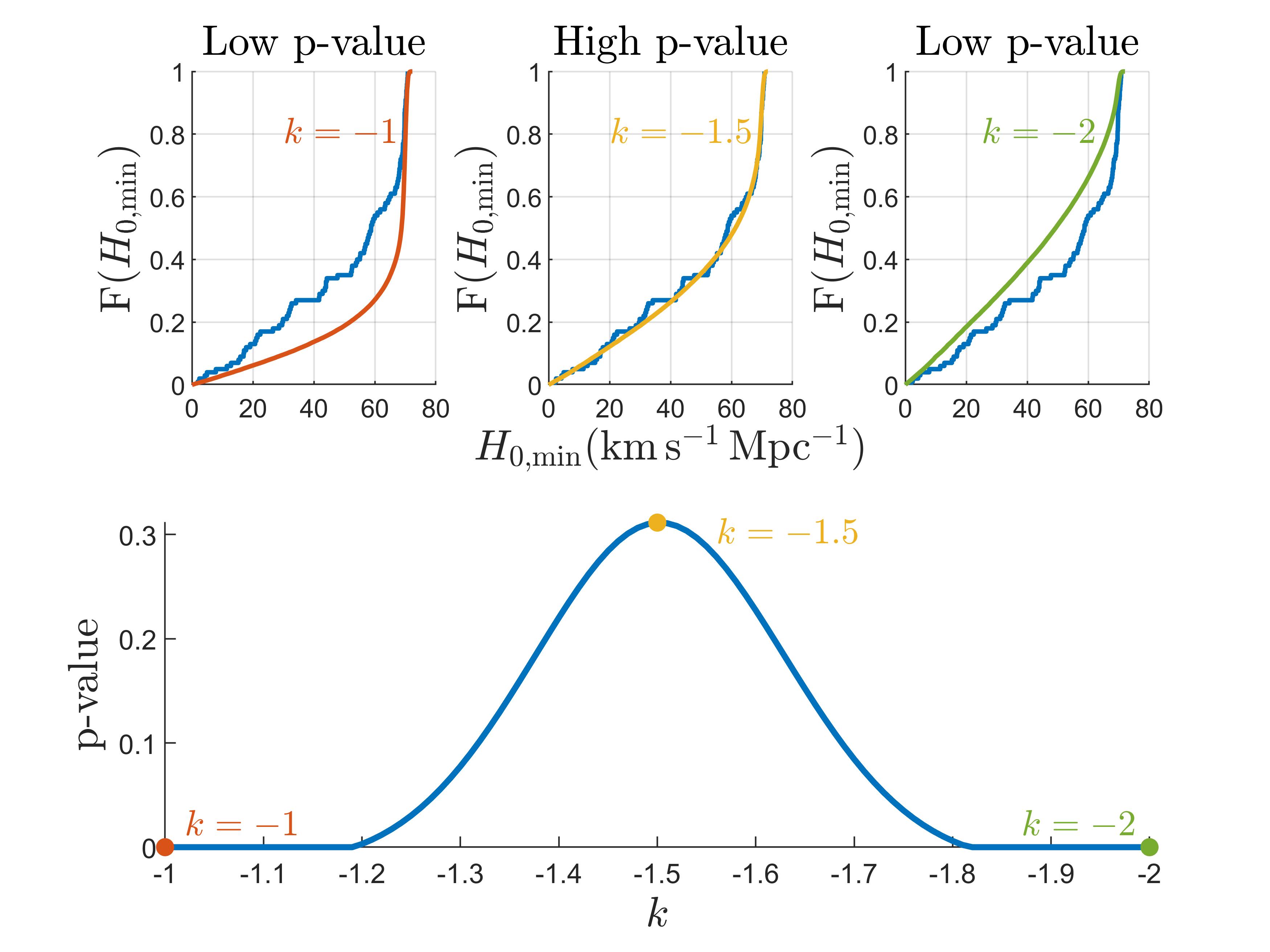}
    	\caption{{Similar to the Fig. \ref{methodH0} but as a function of $k$.}}
    	
        \label{methodk}
        
    \end{figure*}

        \begin{figure*}
    	\includegraphics[width=\textwidth]{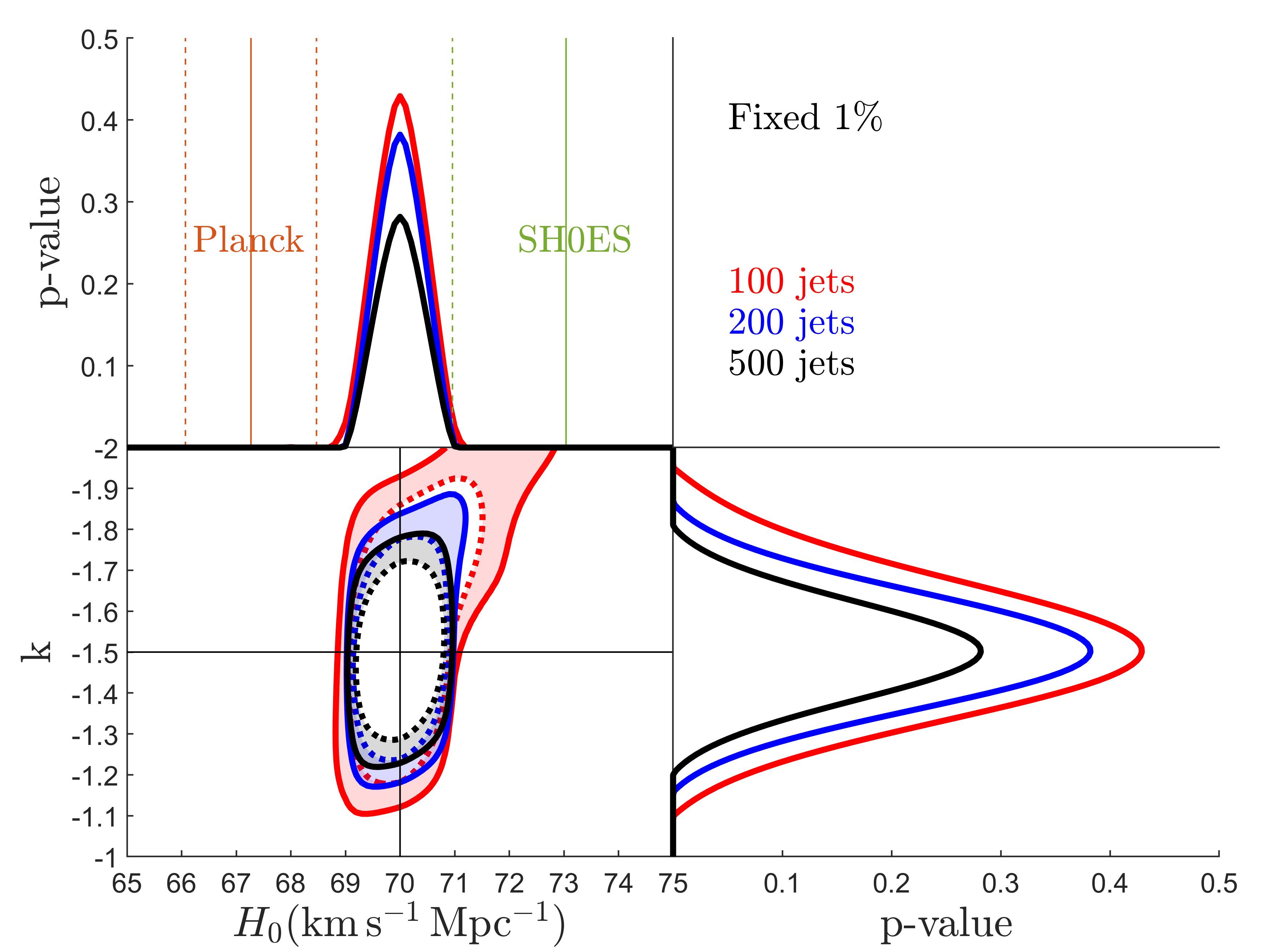}
    	\caption{Constraints of $H_{0}$ and $k$.
    	Three contours show constraints with increasing samples under the same accuracy ($1\%$). Outer contours (solid lines) are 0.01 significance while inner contours (dotted lines) are 0.05 significance  for each scenario. Planck and SH0ES measurements are shown in $2\sigma$.}
    	
        \label{figkfree}
        
    \end{figure*}
    
        \begin{figure*}
    	\includegraphics[width=\textwidth]{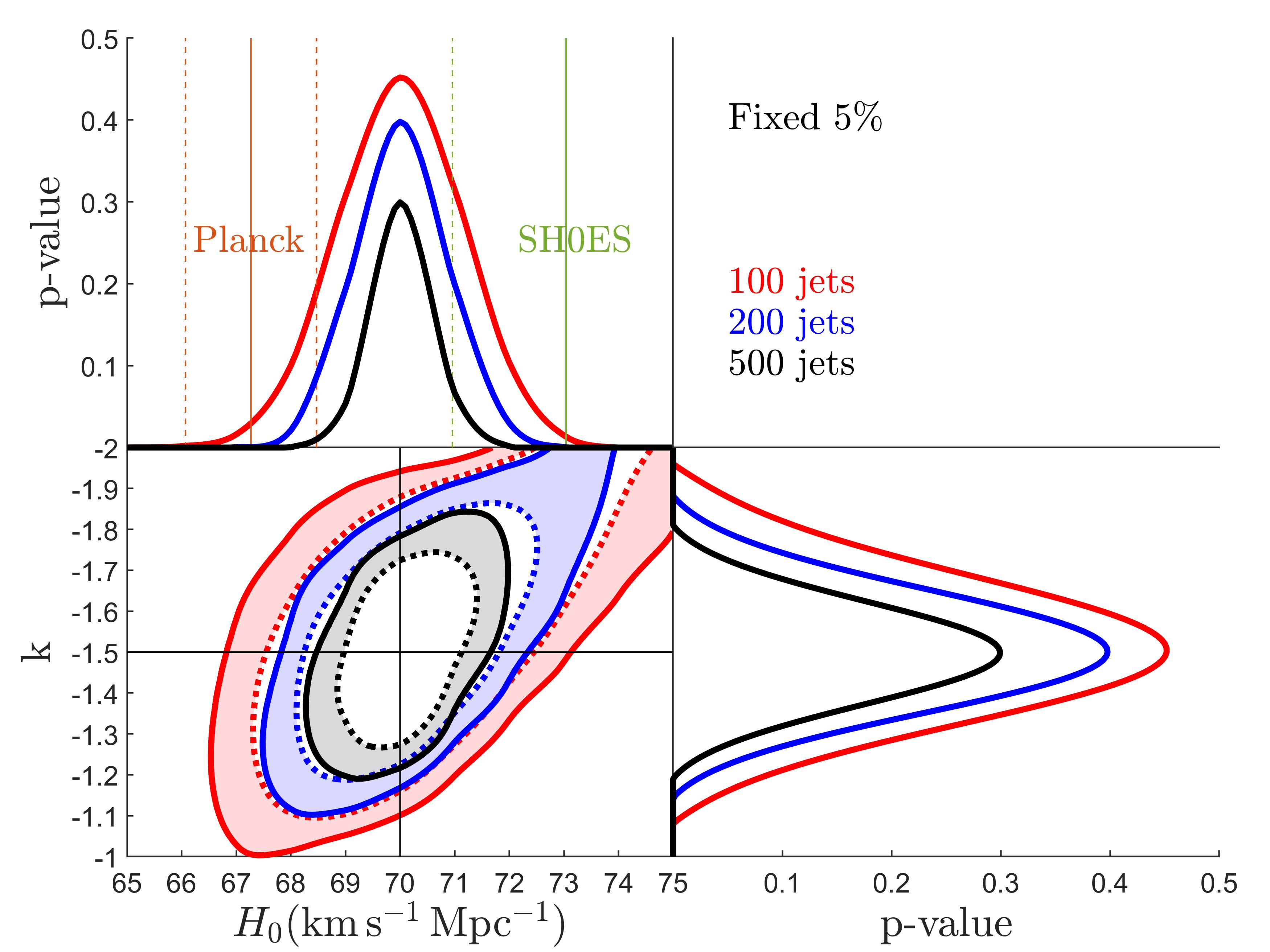}
    	\caption{Same as the Fig. \ref{figkfree} but the accuracy is fixed to $5\%$.}
    	
        \label{figkfree3}
    \end{figure*}    

        \begin{figure*}
    	\includegraphics[width=\textwidth]{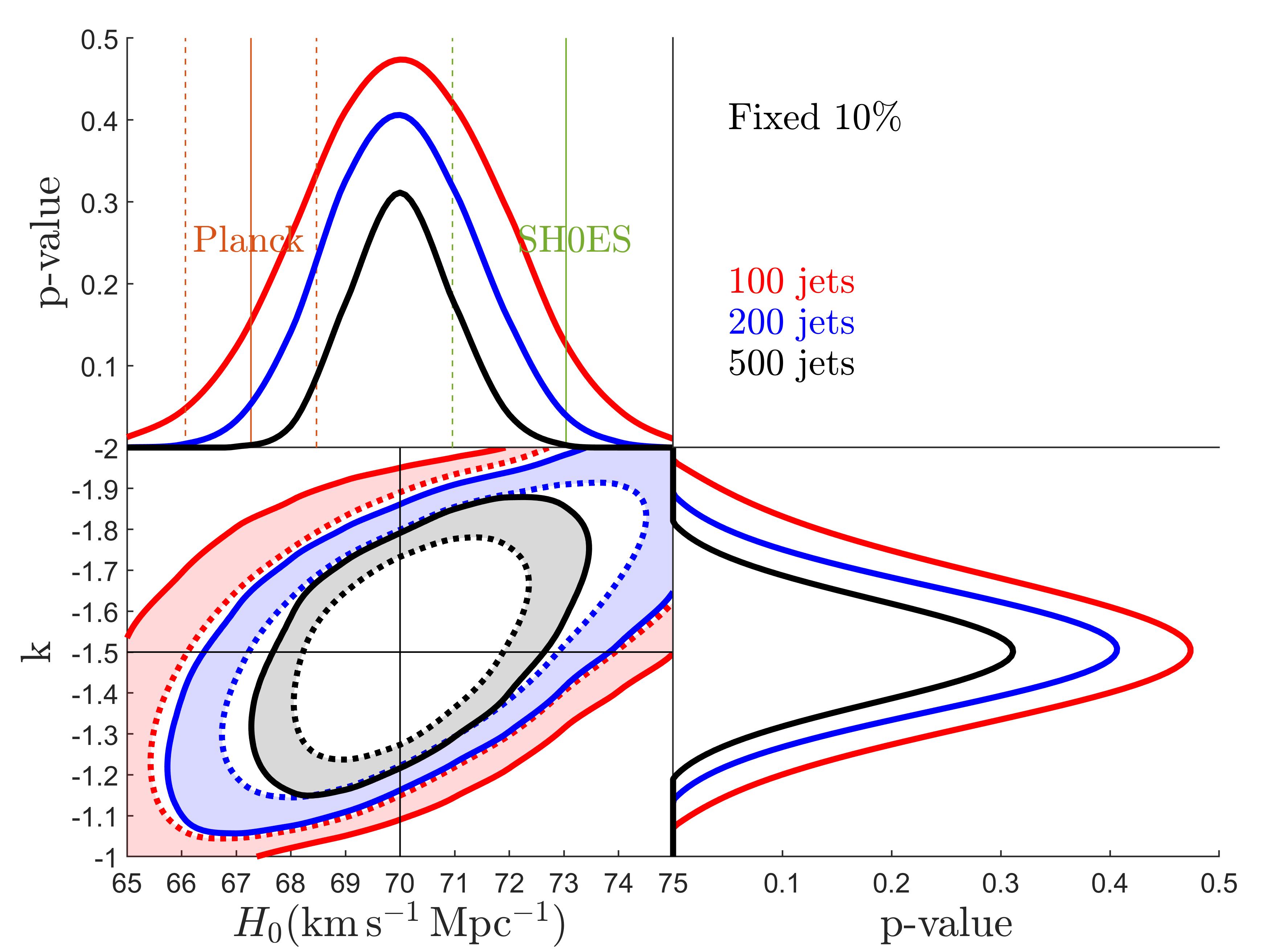}
    	\caption{Same as the Fig. \ref{figkfree} but the accuracy is fixed to $10\%$.}
    	
        \label{figkfree5}
    \end{figure*}        
    
        \begin{figure*}
    	\includegraphics[width=\textwidth]{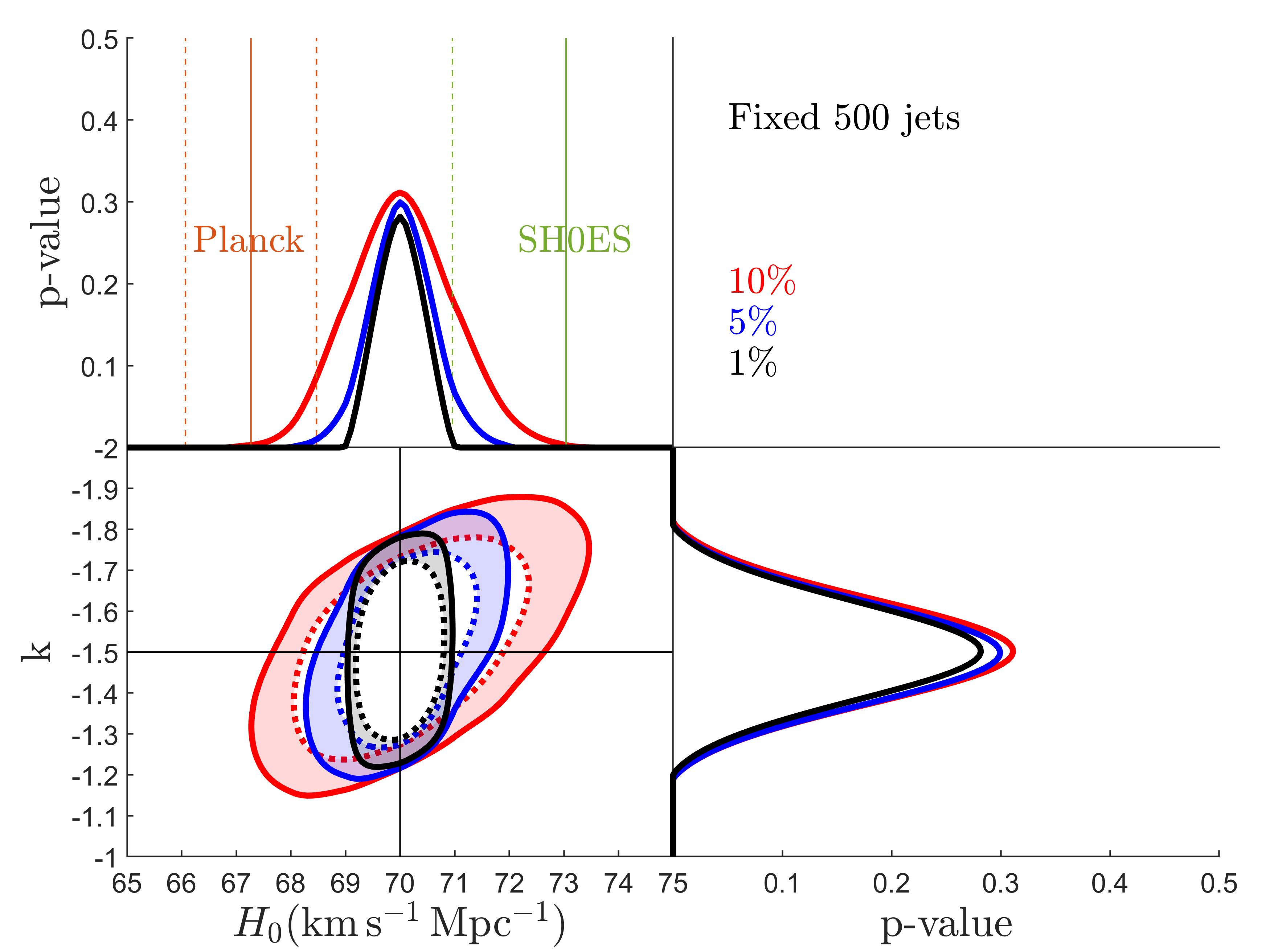}
    	\caption{Constraints of $H_{0}$ and $k$.
    	Three contours show constraints with improving accuracy under the same sample size ($500$ jets). Outer contours (solid lines) are 0.01 significance while inner contours (dotted lines) are 0.05 significance  for each scenario. Planck and SH0ES measurements are shown in $2\sigma$.}
    	
        \label{figkfree2}
    \end{figure*}

        \begin{figure*}
    	\includegraphics[width=\textwidth]{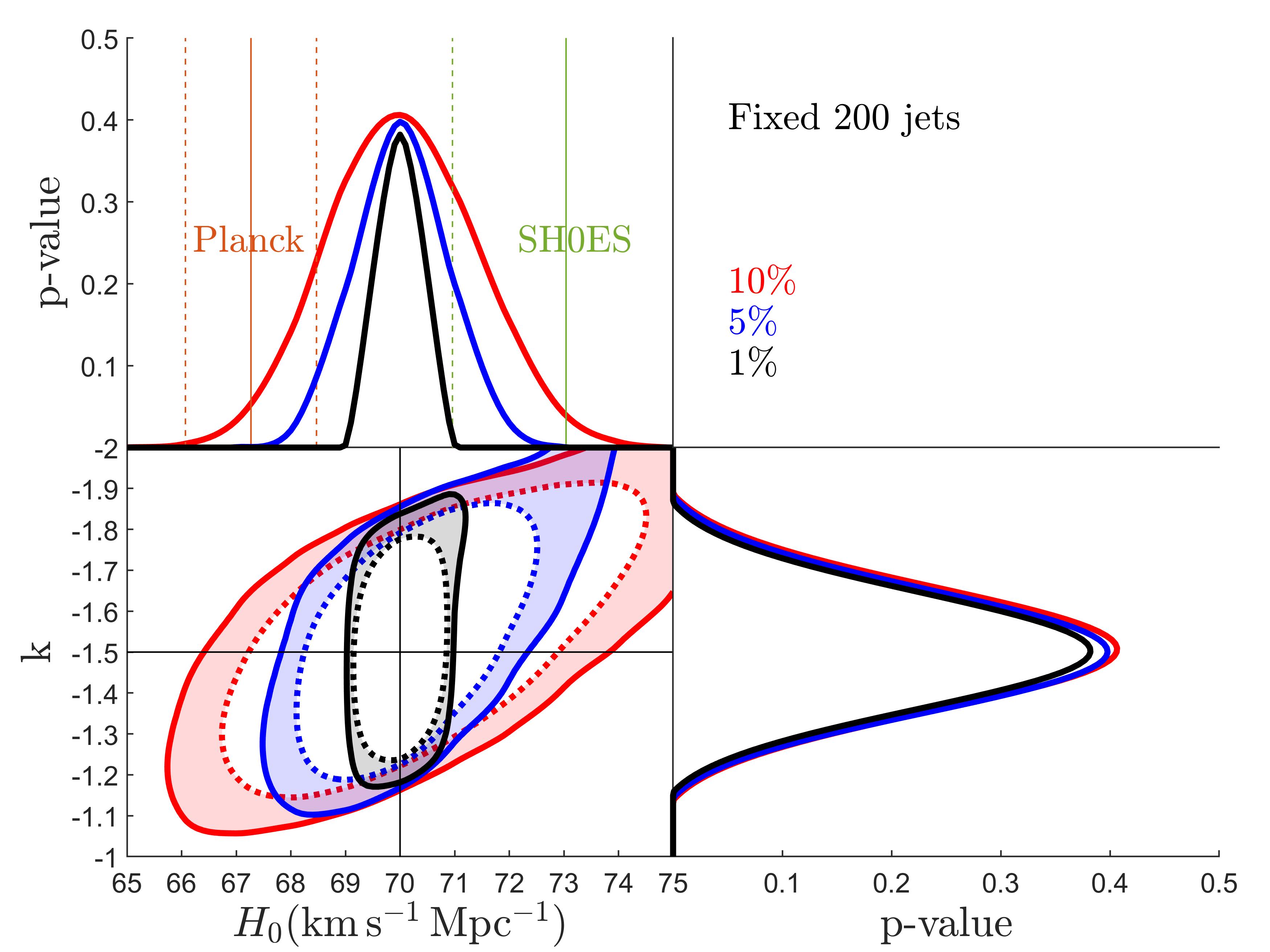}
    	\caption{Same as the Fig. \ref{figkfree2} but the sample size is fixed to 200 jets.}
    	
        \label{figkfree4}
    \end{figure*}

        \begin{figure*}
    	\includegraphics[width=\textwidth]{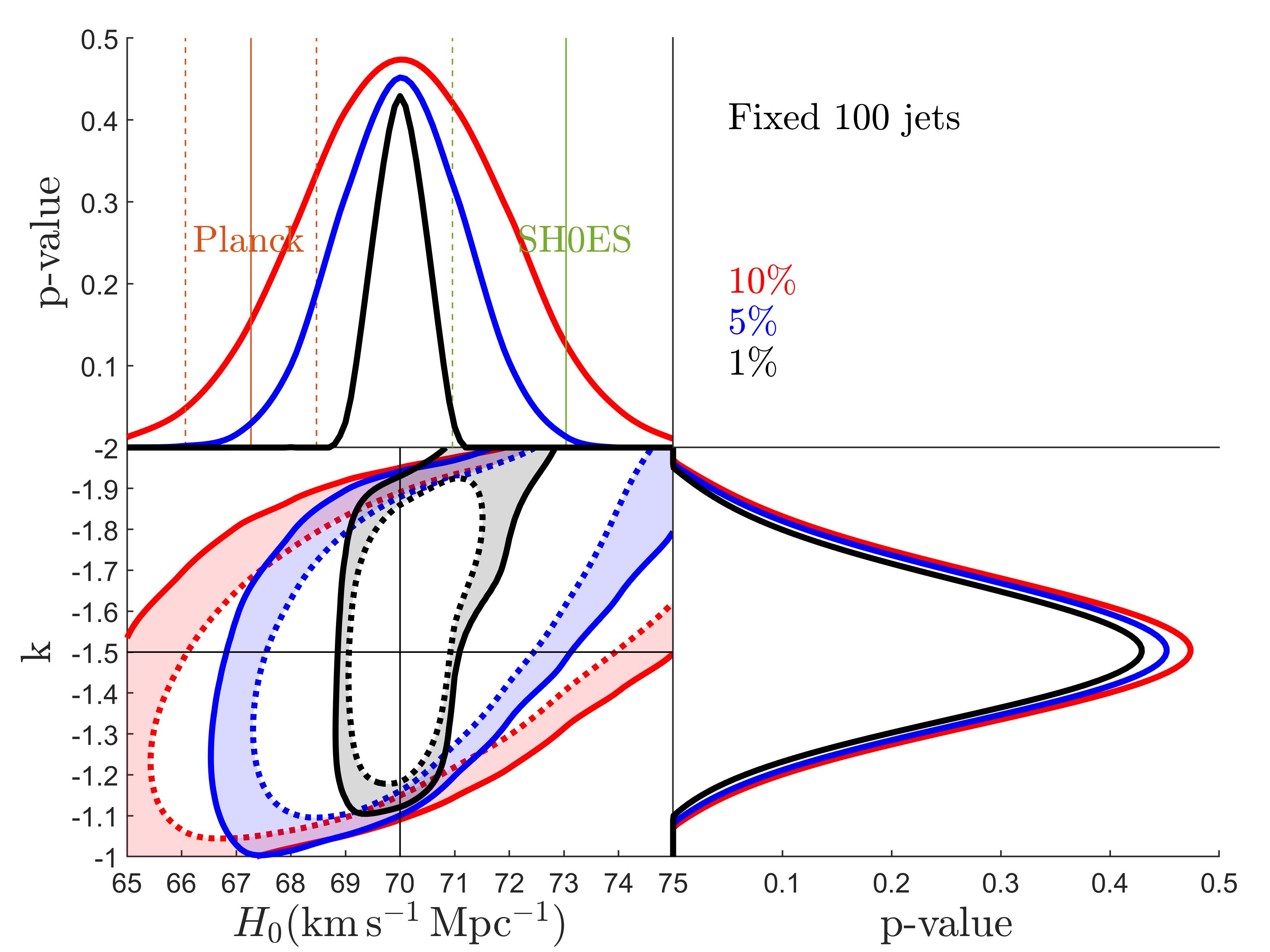}
    	\caption{Same as the Fig. \ref{figkfree2} but the sample size is fixed to 100 jets.}
    	
        \label{figkfree6}
    \end{figure*}


Aside from the highest values of $H_{\rm 0,min}$, a comparison of the whole distribution between theoretical and mock data is also meaningful.
Hence, we propose to perform the K-S test to confirm if the two distributions are statistically different.
By assuming different cosmology and simulating a large enough number of jets (i.e., $10^{5}$), we perform K-S tests between the mock observational data with the theoretical CDF.
Figs. \ref{figkfree} to \ref{figkfree6} show examples of the p-values as a function of $H_{0}$ and $k$ from K-S tests between mock observational data and theoretical profile under different configurations.
{As expected, all of the constraints are centred on the value we assume for the mock observational data ($H_{0}=70\,{\rm km\,s^{-1}\,Mpc^{-1}}$ and $k=-1$).}
Figs. \ref{figkfree}, \ref{figkfree3}, and \ref{figkfree5} demonstrate the constraints of increasing sample sizes under a fixed accuracy (i.e., $1\%$, $5\%$, and $10\%$, respectively).
Figs. \ref{figkfree2}, \ref{figkfree4}, and \ref{figkfree6} demonstrate the constraints of improving accuracy under a fixed sample size (i.e., 500, 200, and 100 jets, respectively).
Our result indicates that when $k$ and $H_{0}$ are close to the real physics in the universe, p-values are higher.
However, there is a degeneracy when the accuracy of jets is high.
Figs. \ref{figkfree} to \ref{figkfree5} demonstrates that if we increase sample sizes, the constraint becomes tighter while the degeneracy remains.
Interestingly, we also find that increasing sample sizes leads to tighter constraints on both {power-law index} and the Hubble constant at moderate accuracy (i.e., $10\%$ and $5\%$) while at $1\%$ accuracy, increasing sample sizes leads to tighter constraints on {power-law index} more. According to Figs. \ref{figkfree2} to \ref{figkfree6}, improving accuracy results in better constraints in the Hubble constant compared with the {power-law index} in all cases but it also alleviates the degeneracy.

{In Fig. \ref{error}, we show the expected error to be constrained of $H_{0}$ (the upper panel) and $k$ (the lower panel) under different sample sizes and the accuracy of the measurement.
Compared with the uncertainties from the SH0ES \citep{Riess2021} and Planck \citep{Planck2020} measurements, our result indicates that if we can meet the criteria of $1\%$ accuracy, the error of $H_{0}$ constrained from the K-S tests will be more accurate.
It is also clear that at $10\%$ and $5\%$ uncertainties, increasing sample sizes will lead to tighter constraints.
However, the constraint saturates at $1\%$ accuracy, which suggests increasing the sample size cannot improve the uncertainties as good as those at $10\%$ and $5\%$.
As can be seen in Fig. \ref{error}, $H_{0}$ can be determined within $1\%$ with 500 jets with $1\%$ accuracy.
Besides, increasing the sample size alone to 500 jets with $10\%$ accuracy meets the precision at the SH0ES level.
As for the power-law index $k$, increasing sample sizes lead to tighter error compared with improving the accuracy of the proper motion.}

{We also note that due to the Doppler beaming effect (see the details in the Sec. \ref{Method}), many receding jets are not easily observable.
In our simulations, we simply assume that all jets among our samples have measurable receding jets.
The Doppler de-boosting effect makes the receding jets dimmer.
Improving the integration time and the sensitivity of the telescope will help us to address this issue in the future.
Besides, our method requires a large enough sample size to have a meaningful statistical result from the K-S test.
For instance, in the MOJAVE program, there are $\sim90$ jets that are two-sided out of $\sim500$ samples.
Many of the samples are only detected one-sided due to the Doppler boosting effect.
We suggest three ways to increase the sample size.
The first way is to increase the total sample of radio sources.
On-going and future radio sky surveys such as the Very Large Array Sky Survey \citep[VLASS;][]{Lacy2020} which is expected to detect 10 million radio sources including radio galaxies (with two-sided jets).
The second way is to improve the sensitivity of the telescope.
Once the limiting magnitude reaches the brightness of the receding jets, we will be able to measure the proper motions from both jets.
For example, for an approaching jet with $\gamma=3$ and $\theta=\pi/6$, the Doppler factor is $\sim1.82$.
Therefore, the Doppler factor of the receding jet is expected to be $\sim0.18$.
Adopting $p-\alpha=3$ in the Eq. \ref{dopplerbeaming}, the receding jet is $\sim1000$ times fainter than the approaching jet due to the Doppler deboosting effect (and boosting effect for the approaching jet).
If the SNR of the approaching jet is 1000 \citep[e.g.,][]{Lister2019,Baczko2019}, improving the sensitivity to $\sim3$ to $\sim5$ times better will be able to detect it.
\citet{Hovatta2009} calculated the Doppler factors of quasars, BL Lacertae objects, and radio galaxies.
The sources with low-Doppler factors are worth further observation on the receding jets when improving sensitivity in the near future.
The third way is to monitor the two-sided jets which have not been monitored before.
For instance, in the catalogue of FR I radio galaxies (with 219 FR I radio galaxies), FRICAT \citep{Capetti2017}, many jets are two-sided but only have image data, which suggests those jets have no measurement of proper motions.}

\begin{figure}
    	\includegraphics[width=\columnwidth]{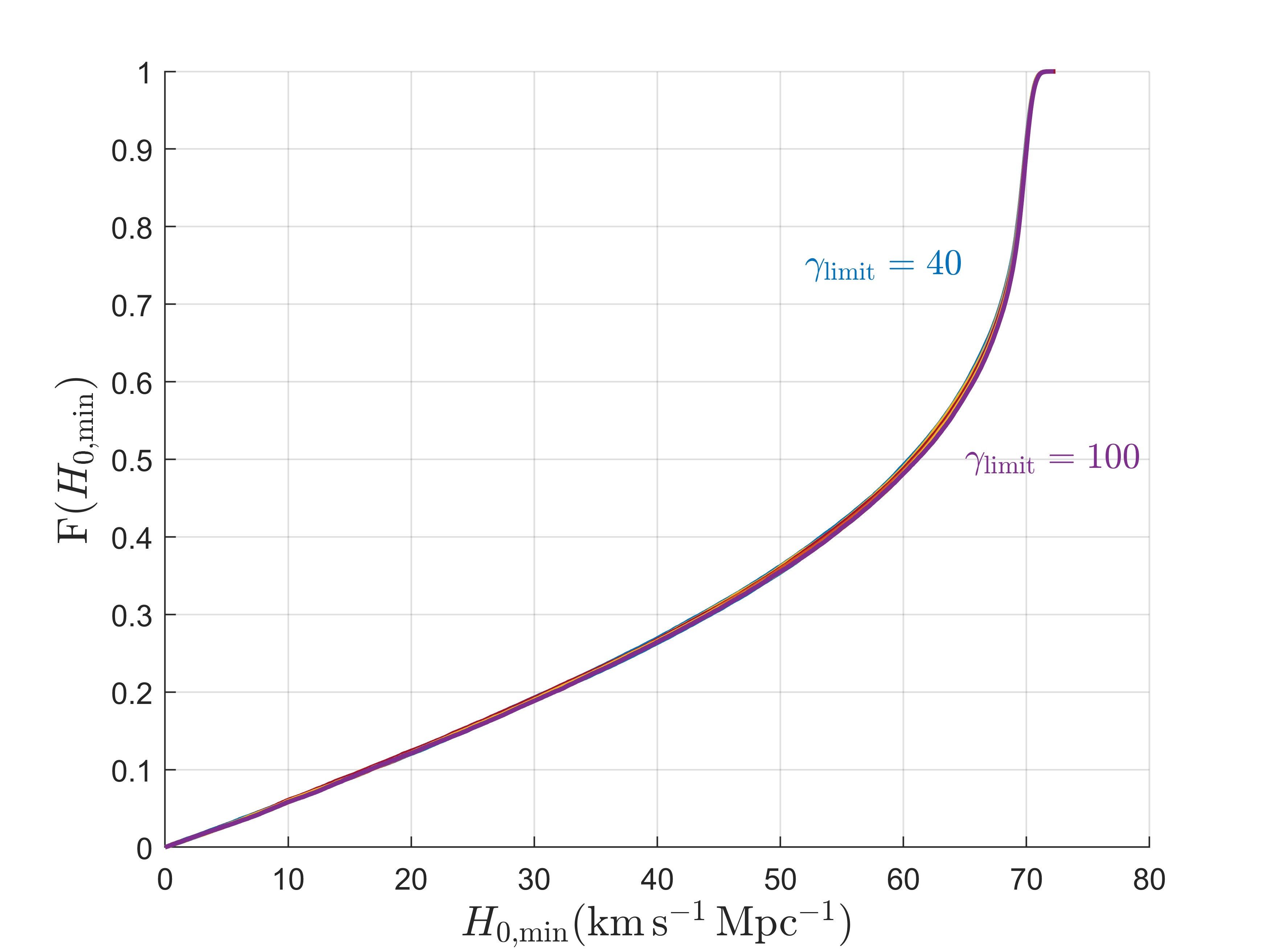}
    	\caption{{Similar to Fig. \ref{theo}. CDFs of $H_{\rm 0,min}$ under different prior $\gamma_{\rm upper}$.}}
    	
        \label{theory_gamma}
        
    \end{figure}

\begin{figure}
    	\includegraphics[width=\columnwidth]{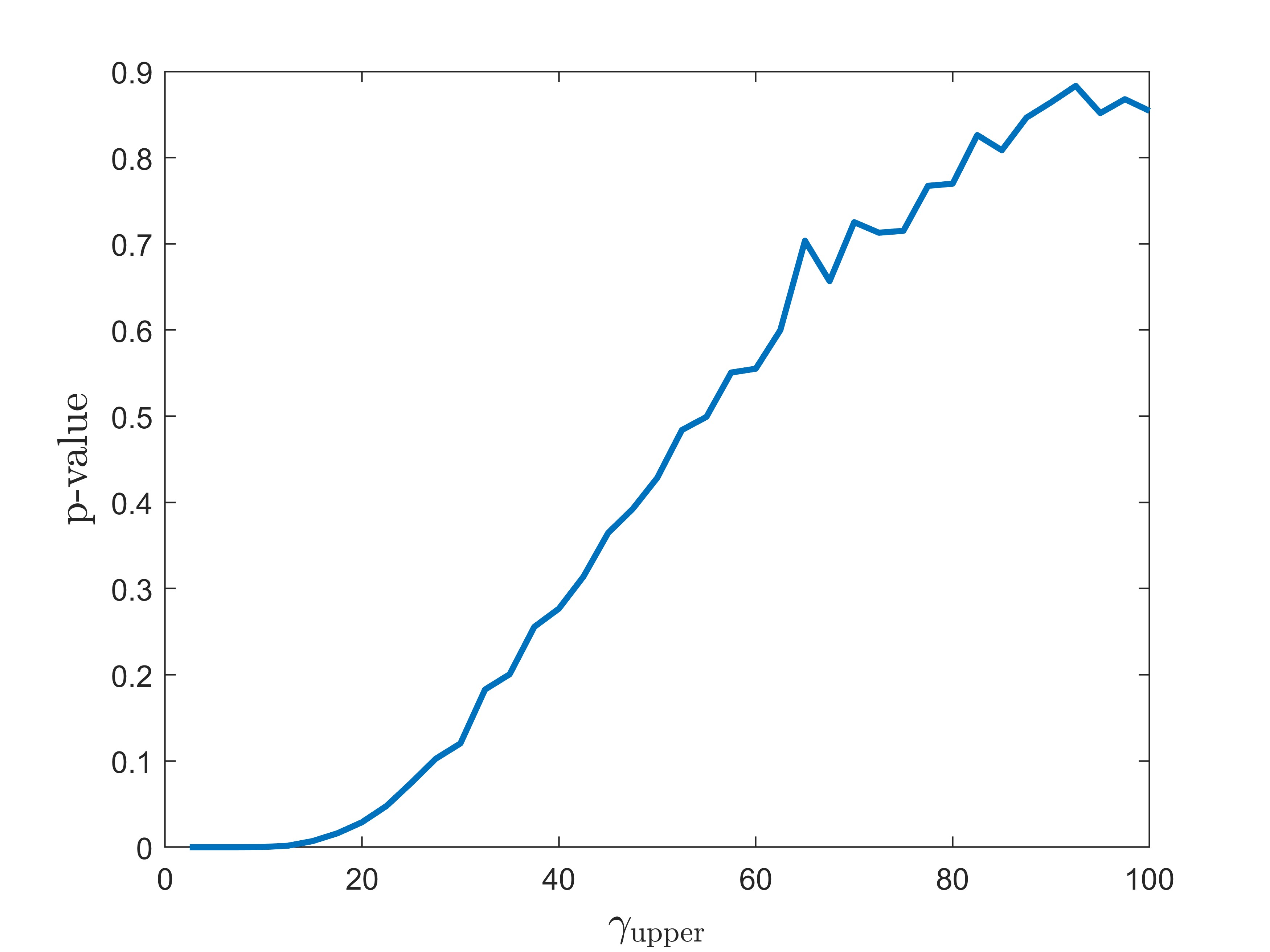}
    	\caption{{The p-value as a function of $\gamma_{\rm upper}$. The mock observational data is same as the data presented in Fig. \ref{methodH0}}.}
    	
        \label{p_gamma}
        
    \end{figure}

\begin{figure}
    	\includegraphics[width=\columnwidth]{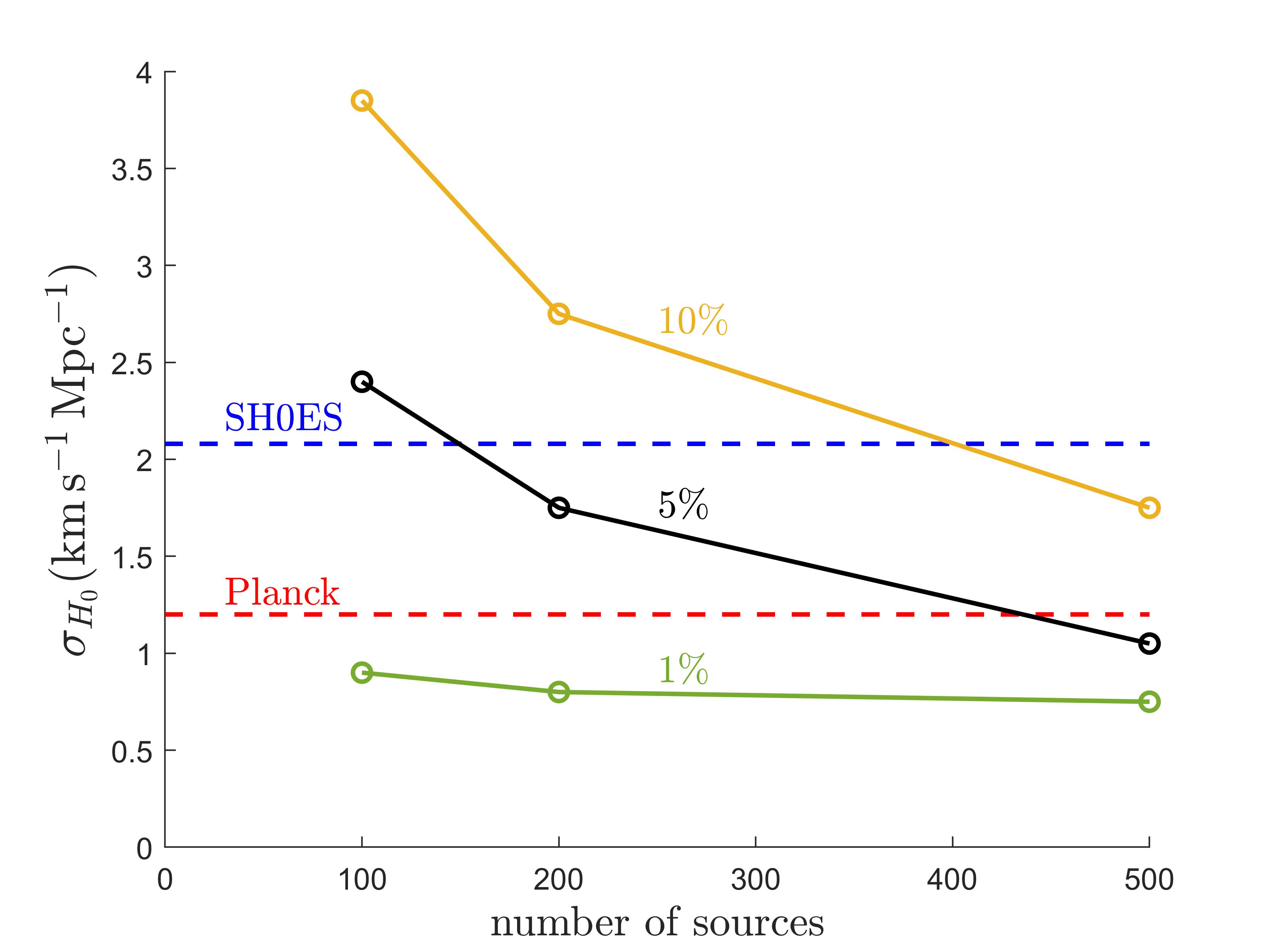}
    	\includegraphics[width=\columnwidth]{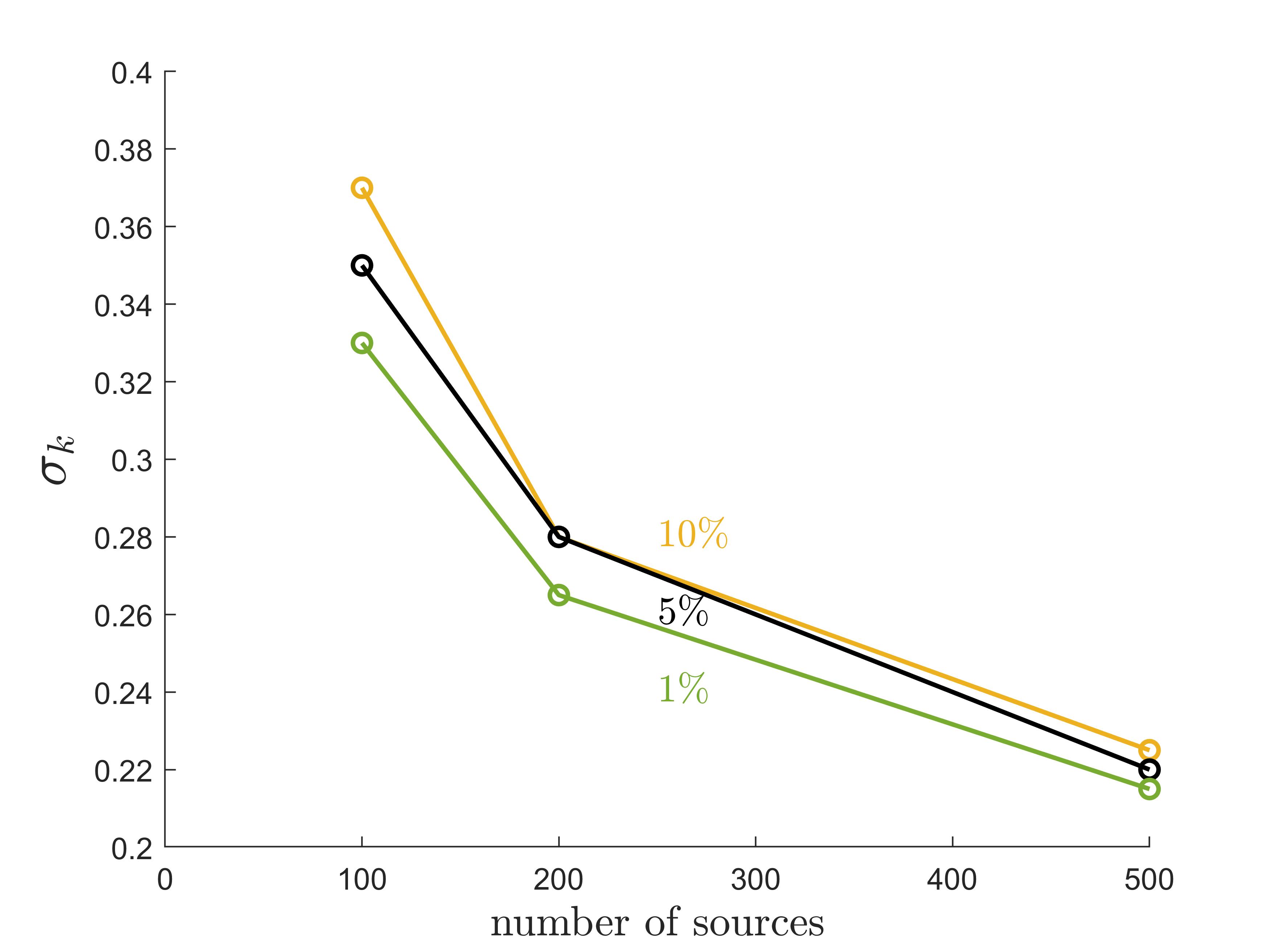}
    	\caption{{The errors (p-value$>0.05$) under different configurations. The upper panel shows the errors of $H_{0}$ with different sample sizes and different accuracies compared with the previous error ($2\sigma$ are shown to compare fairly with our p-value$=0.05$) of the $H_{0}$ measurement of SH0ES \citep{Riess2021} and Planck \citep{Planck2020}. The lower panel shows the errors of $k$ under different configurations.}}
    	
        \label{error}
        
    \end{figure}

{In addition, it might be difficult to collect many proper motions of twin-jets with high Lorentz factor, especially the receding components.
Therefore, we also test if the method still works under different maximum values of the velocity ($\gamma_{\rm upper}$).
In our mock observational data, we adopt $\gamma_{\rm upper}=100$.
We show how CDF changes under different $\gamma_{\rm upper}$ in Fig. \ref{theory_gamma}.
Regarding $H_{0}$ and $k$, $\gamma_{\rm upper}$ does not affect the CDF significantly.
Here, we gradually reduce the maximum value of the Lorentz factor for the theoretical CDF and conduct K-S tests with the mock observational data.
The assumption of the mock observational data is identical to that in Fig. \ref{methodH0}, with 100 jets and $1\%$ accuracy.
The result is shown in Fig. \ref{p_gamma}, which suggests that the critical value is at $\gamma_{\rm upper}\sim30$.
Values with $\gamma_{\rm upper}>30$ have p-values higher than 0.05.
Once the maximum value of the Lorentz factor is smaller than the critical value, 30, our method fails with p-values smaller than 0.05.
}

{Based on Fig. 2 and Tab. 1  of \citet{Pracy2016}, there are $\sim2200$ radio galaxies at $z<0.3$.
Assuming the ratio of two-sided jets and total jet sample in the MOJAVE program ($\sim90/\sim500\sim18\%$), we expect there are $2200\times18\%\sim400$ two-sided jets in the radio galaxy catalogue.
Conducting VLBA to observe such samples will bring the two-sided jets to 500 including the 90 in the MOJAVE program.
In the MOJAVE program, they monitored 447 AGN jets from 25-year multi-epoch observations \citep{Homan2021}.
Thus, if aiming at these two-sided jets, $\sim20$ years of observation at different epochs to measure the proper motion \citep[e.g., $\sim1\,$mas/yr for a jet of $\beta\sim0.6$ at $z\sim0.005$; $\sim100\,\rm\mu as/yr$ for a jet of $\beta\sim0.9$ at $z\sim0.3$;][]{Baczko2019} with VLBA ($\sim$mas resolution) may increase the sample size of two-sided jets to $\sim500$ in the local Universe.
The uncertainties of the proper motions reach $\rm \sim5 \mu as$ level with integration times of $30$ to $50$ minutes \citep[e.g.,][]{Lister2019}.
Also, the radio galaxies included in \citet{Pracy2016} have a criteria with flux density at $1.4\,{\rm GHz}>2.8\,{\rm mJy}$.
Therefore, MOJAVE VLBA with image rms levels of $\sim0.1\,{\rm mJy\,beam^{-1}}$ is promising to detect and monitor these radio galaxies.}

{
Last but not least, if there are too few samples in the nearby Universe, we can extend to higher redshift as long as the approximation of the FLRW metric still works.
Moreover, even if the approximation fails at high redshift, if we assume cosmological parameters other than $H_{0}$, we can extend this method to higher redshifts.
For instance, for a $\beta=0.9999$ ($\gamma\sim70$), $\theta=\pi/2$, the $H_{0}$ is only $\sim1\%$ more than the assumed $H_{0}$ at $z=0.05$.
We can also revise the redshift distribution based on the distribution from observation for theoretical (simulated) CDFs.}

\section{Conclusion}
\label{Conclusion}
Based on FLRW cosmology and geometrical relation of the proper motion, the lower limit ($H_{\rm 0,min}$) of the Hubble constant ($H_{0}$) can be determined only with three cosmology-free observables: redshift, approaching proper motion, and receding proper motion.
We propose a new method, performing K-S tests between the observational (mock) data and the theoretical CDF of different $H_{0}$ and $k$, in order to constrain $H_{0}$ and $k$.
We simulate $10^{5}$ jets as a numerical distribution of $H_{\rm 0,min}$ and $k$ between $65\,{\rm km\,s^{-1}\,Mpc^{-1}}<H_{0}<75\,{\rm km\,s^{-1}\,Mpc^{-1}}$ and $-1<k<-2$, as theoretical distributions and create mock observational data.
The result shows that we can simultaneously constrain $H_{0}$ and $k$.
If the values of $H_{0}$ and $k$ are close to the real universe, p-values are higher.
We find that there is a degeneracy between $H_{0}$ and $k$ when the accuracy is high (e.g., $10\%$).
Interestingly, we also find that increasing sample sizes leads to tighter constraints on both {power-law index} and the Hubble constant at moderate accuracy (i.e., $10\%$ and $5\%$) while at $1\%$ accuracy, increasing sample sizes leads to tighter constraints on {power-law index} more. Improving accuracy results in better constraints in the Hubble constant compared with the {power-law index} in all cases but it also alleviates the degeneracy.
In the future, we will be able to alleviate the Hubble tension by the distribution of $H_{\rm 0,min}$ calculated from the proper motion and we could also further constrain $H_{0}$ and $k$.

\section*{Acknowledgements}
{We thank the anonymous referee for useful comments and constructive remarks on the manuscript.}
TG acknowledges the support of the National Science and Technology Council of Taiwan through grants 108-2628-M-007-004-MY3 and 111-2123-M-001-008-.
TH acknowledges the support of the National Science and Technology Council of Taiwan through grants 110-2112-M-005-013-MY3, 110-2112-M-007-034-, and 111-2123-M-001-008-.

\section*{Data Availability}
\label{Data availability}
The data underlying this article will be shared on reasonable request to the corresponding author.

\bibliographystyle{mnras}
\bibliography{references}

\begin{thebibliography}{}
\makeatletter
\relax
\def\mn@urlcharsother{\let\do\@makeother \do\$\do\&\do\#\do\^\do\_\do\%\do\~}
\def\mn@doi{\begingroup\mn@urlcharsother \@ifnextchar [ {\mn@doi@}
  {\mn@doi@[]}}
\def\mn@doi@[#1]#2{\def\@tempa{#1}\ifx\@tempa\@empty \href
  {http://dx.doi.org/#2} {doi:#2}\else \href {http://dx.doi.org/#2} {#1}\fi
  \endgroup}
\def\mn@eprint#1#2{\mn@eprint@#1:#2::\@nil}
\def\mn@eprint@arXiv#1{\href {http://arxiv.org/abs/#1} {{\tt arXiv:#1}}}
\def\mn@eprint@dblp#1{\href {http://dblp.uni-trier.de/rec/bibtex/#1.xml}
  {dblp:#1}}
\def\mn@eprint@#1:#2:#3:#4\@nil{\def\@tempa {#1}\def\@tempb {#2}\def\@tempc
  {#3}\ifx \@tempc \@empty \let \@tempc \@tempb \let \@tempb \@tempa \fi \ifx
  \@tempb \@empty \def\@tempb {arXiv}\fi \@ifundefined
  {mn@eprint@\@tempb}{\@tempb:\@tempc}{\expandafter \expandafter \csname
  mn@eprint@\@tempb\endcsname \expandafter{\@tempc}}}

\bibitem[\protect\citeauthoryear{{Abbott} et~al.,}{{Abbott}
  et~al.}{2017}]{Abbott2017}
{Abbott} B.~P.,  et~al., 2017, \mn@doi [\nat] {10.1038/nature24471}, \href
  {https://ui.adsabs.harvard.edu/abs/2017Natur.551...85A} {551, 85}

\bibitem[\protect\citeauthoryear{{Adhikari} et~al.,}{{Adhikari}
  et~al.}{2017}]{Adhikari2017}
{Adhikari} R.,  et~al., 2017, \mn@doi [\jcap] {10.1088/1475-7516/2017/01/025},
  \href {https://ui.adsabs.harvard.edu/abs/2017JCAP...01..025A} {2017, 025}

\bibitem[\protect\citeauthoryear{{Ajello} et~al.,}{{Ajello}
  et~al.}{2012}]{Ajello2012}
{Ajello} M.,  et~al., 2012, \mn@doi [\apj] {10.1088/0004-637X/751/2/108}, \href
  {https://ui.adsabs.harvard.edu/abs/2012ApJ...751..108A} {751, 108}

\bibitem[\protect\citeauthoryear{{Baczko}, {Schulz}, {Kadler}, {Ros},
  {Perucho}, {Fromm}  \& {Wilms}}{{Baczko} et~al.}{2019}]{Baczko2019}
{Baczko} A.~K.,  {Schulz} R.,  {Kadler} M.,  {Ros} E.,  {Perucho} M.,  {Fromm}
  C.~M.,   {Wilms} J.,  2019, \mn@doi [\aap] {10.1051/0004-6361/201833828},
  \href {https://ui.adsabs.harvard.edu/abs/2019A&A...623A..27B} {623, A27}

\bibitem[\protect\citeauthoryear{{Behr}, {Schucking}, {Vishveshwara}  \&
  {Wallace}}{{Behr} et~al.}{1976}]{Behr1976}
{Behr} C.,  {Schucking} E.~L.,  {Vishveshwara} C.~V.,   {Wallace} W.,  1976,
  \mn@doi [\aj] {10.1086/111869}, \href
  {https://ui.adsabs.harvard.edu/abs/1976AJ.....81..147B} {81, 147}

\bibitem[\protect\citeauthoryear{{Blandford} \& {K{\"o}nigl}}{{Blandford} \&
  {K{\"o}nigl}}{1979}]{Blandford1979}
{Blandford} R.~D.,  {K{\"o}nigl} A.,  1979, \mn@doi [\apj] {10.1086/157262},
  \href {https://ui.adsabs.harvard.edu/abs/1979ApJ...232...34B} {232, 34}

\bibitem[\protect\citeauthoryear{{Blandford} \& {Znajek}}{{Blandford} \&
  {Znajek}}{1977}]{Blandford1977}
{Blandford} R.~D.,  {Znajek} R.~L.,  1977, \mn@doi [\mnras]
  {10.1093/mnras/179.3.433}, \href
  {https://ui.adsabs.harvard.edu/abs/1977MNRAS.179..433B} {179, 433}

\bibitem[\protect\citeauthoryear{{Blandford}, {McKee}  \& {Rees}}{{Blandford}
  et~al.}{1977}]{Blandford1977b}
{Blandford} R.~D.,  {McKee} C.~F.,   {Rees} M.~J.,  1977, \mn@doi [\nat]
  {10.1038/267211a0}, \href
  {https://ui.adsabs.harvard.edu/abs/1977Natur.267..211B} {267, 211}

\bibitem[\protect\citeauthoryear{{Bridle} \& {Perley}}{{Bridle} \&
  {Perley}}{1984}]{Bridle1984}
{Bridle} A.~H.,  {Perley} R.~A.,  1984, \mn@doi [\araa]
  {10.1146/annurev.aa.22.090184.001535}, \href
  {https://ui.adsabs.harvard.edu/abs/1984ARA&A..22..319B} {22, 319}

\bibitem[\protect\citeauthoryear{{Capetti}, {Massaro}  \& {Baldi}}{{Capetti}
  et~al.}{2017}]{Capetti2017}
{Capetti} A.,  {Massaro} F.,   {Baldi} R.~D.,  2017, \mn@doi [\aap]
  {10.1051/0004-6361/201629287}, \href
  {https://ui.adsabs.harvard.edu/abs/2017A&A...598A..49C} {598, A49}

\bibitem[\protect\citeauthoryear{{Cara} \& {Lister}}{{Cara} \&
  {Lister}}{2008}]{Cara2008}
{Cara} M.,  {Lister} M.~L.,  2008, \mn@doi [\apj] {10.1086/525554}, \href
  {https://ui.adsabs.harvard.edu/abs/2008ApJ...674..111C} {674, 111}

\bibitem[\protect\citeauthoryear{{Chen}, {Fishbach}  \& {Holz}}{{Chen}
  et~al.}{2018}]{Chen2018}
{Chen} H.-Y.,  {Fishbach} M.,   {Holz} D.~E.,  2018, \mn@doi [\nat]
  {10.1038/s41586-018-0606-0}, \href
  {https://ui.adsabs.harvard.edu/abs/2018Natur.562..545C} {562, 545}

\bibitem[\protect\citeauthoryear{{Cohen}, {Lister}, {Homan}, {Kadler},
  {Kellermann}, {Kovalev}  \& {Vermeulen}}{{Cohen} et~al.}{2007}]{Cohen2007}
{Cohen} M.~H.,  {Lister} M.~L.,  {Homan} D.~C.,  {Kadler} M.,  {Kellermann}
  K.~I.,  {Kovalev} Y.~Y.,   {Vermeulen} R.~C.,  2007, \mn@doi [\apj]
  {10.1086/511063}, \href
  {https://ui.adsabs.harvard.edu/abs/2007ApJ...658..232C} {658, 232}

\bibitem[\protect\citeauthoryear{{Cuceu}, {Farr}, {Lemos}  \&
  {Font-Ribera}}{{Cuceu} et~al.}{2019}]{Cuceu2019}
{Cuceu} A.,  {Farr} J.,  {Lemos} P.,   {Font-Ribera} A.,  2019, \mn@doi [\jcap]
  {10.1088/1475-7516/2019/10/044}, \href
  {https://ui.adsabs.harvard.edu/abs/2019JCAP...10..044C} {2019, 044}

\bibitem[\protect\citeauthoryear{{Eisenstein} et~al.,}{{Eisenstein}
  et~al.}{2005}]{Eisenstein2005}
{Eisenstein} D.~J.,  et~al., 2005, \mn@doi [\apj] {10.1086/466512}, \href
  {https://ui.adsabs.harvard.edu/abs/2005ApJ...633..560E} {633, 560}

\bibitem[\protect\citeauthoryear{{Hashimoto}, {Goto}, {Wang}, {Kim}, {Wu}  \&
  {Ho}}{{Hashimoto} et~al.}{2019}]{Hashimoto2019}
{Hashimoto} T.,  {Goto} T.,  {Wang} T.-W.,  {Kim} S.~J.,  {Wu} Y.-H.,   {Ho}
  C.-C.,  2019, \mn@doi [\mnras] {10.1093/mnras/stz1715}, \href
  {https://ui.adsabs.harvard.edu/abs/2019MNRAS.488.1908H} {488, 1908}

\bibitem[\protect\citeauthoryear{{Hawley} \& {Balbus}}{{Hawley} \&
  {Balbus}}{2002}]{Hawley2002}
{Hawley} J.~F.,  {Balbus} S.~A.,  2002, \mn@doi [\apj] {10.1086/340765}, \href
  {https://ui.adsabs.harvard.edu/abs/2002ApJ...573..738H} {573, 738}

\bibitem[\protect\citeauthoryear{{Herrnstein} et~al.,}{{Herrnstein}
  et~al.}{1999}]{Herrnstein1999}
{Herrnstein} J.~R.,  et~al., 1999, \mn@doi [\nat] {10.1038/22972}, \href
  {https://ui.adsabs.harvard.edu/abs/1999Natur.400..539H} {400, 539}

\bibitem[\protect\citeauthoryear{{Homan} et~al.,}{{Homan}
  et~al.}{2021}]{Homan2021}
{Homan} D.~C.,  et~al., 2021, \mn@doi [\apj] {10.3847/1538-4357/ac27af}, \href
  {https://ui.adsabs.harvard.edu/abs/2021ApJ...923...67H} {923, 67}

\bibitem[\protect\citeauthoryear{{Hotokezaka}, {Nakar}, {Gottlieb}, {Nissanke},
  {Masuda}, {Hallinan}, {Mooley}  \& {Deller}}{{Hotokezaka}
  et~al.}{2019}]{Hotokezaka2019}
{Hotokezaka} K.,  {Nakar} E.,  {Gottlieb} O.,  {Nissanke} S.,  {Masuda} K.,
  {Hallinan} G.,  {Mooley} K.~P.,   {Deller} A.~T.,  2019, \mn@doi [Nature
  Astronomy] {10.1038/s41550-019-0820-1}, \href
  {https://ui.adsabs.harvard.edu/abs/2019NatAs...3..940H} {3, 940}

\bibitem[\protect\citeauthoryear{{Hovatta}, {Valtaoja}, {Tornikoski}  \&
  {L{\"a}hteenm{\"a}ki}}{{Hovatta} et~al.}{2009}]{Hovatta2009}
{Hovatta} T.,  {Valtaoja} E.,  {Tornikoski} M.,   {L{\"a}hteenm{\"a}ki} A.,
  2009, \mn@doi [\aap] {10.1051/0004-6361:200811150}, \href
  {https://ui.adsabs.harvard.edu/abs/2009A&A...494..527H} {494, 527}

\bibitem[\protect\citeauthoryear{{Hubble}}{{Hubble}}{1929}]{Hubble1929}
{Hubble} E.,  1929, \mn@doi [Proceedings of the National Academy of Science]
  {10.1073/pnas.15.3.168}, \href
  {https://ui.adsabs.harvard.edu/abs/1929PNAS...15..168H} {15, 168}

\bibitem[\protect\citeauthoryear{{Humphreys}, {Reid}, {Moran}, {Greenhill}  \&
  {Argon}}{{Humphreys} et~al.}{2013}]{Humphreys2013}
{Humphreys} E.~M.~L.,  {Reid} M.~J.,  {Moran} J.~M.,  {Greenhill} L.~J.,
  {Argon} A.~L.,  2013, \mn@doi [\apj] {10.1088/0004-637X/775/1/13}, \href
  {https://ui.adsabs.harvard.edu/abs/2013ApJ...775...13H} {775, 13}

\bibitem[\protect\citeauthoryear{{Lacy} et~al.,}{{Lacy}
  et~al.}{2020}]{Lacy2020}
{Lacy} M.,  et~al., 2020, \mn@doi [\pasp] {10.1088/1538-3873/ab63eb}, \href
  {https://ui.adsabs.harvard.edu/abs/2020PASP..132c5001L} {132, 035001}

\bibitem[\protect\citeauthoryear{{Laing} \& {Bridle}}{{Laing} \&
  {Bridle}}{2002}]{Laing2002}
{Laing} R.~A.,  {Bridle} A.~H.,  2002, \mn@doi [\mnras]
  {10.1046/j.1365-8711.2002.05756.x}, \href
  {https://ui.adsabs.harvard.edu/abs/2002MNRAS.336..328L} {336, 328}

\bibitem[\protect\citeauthoryear{{Law}, {Steidel}, {Erb}, {Larkin}, {Pettini},
  {Shapley}  \& {Wright}}{{Law} et~al.}{2009}]{Law2009}
{Law} D.~R.,  {Steidel} C.~C.,  {Erb} D.~K.,  {Larkin} J.~E.,  {Pettini} M.,
  {Shapley} A.~E.,   {Wright} S.~A.,  2009, \mn@doi [\apj]
  {10.1088/0004-637X/697/2/2057}, \href
  {https://ui.adsabs.harvard.edu/abs/2009ApJ...697.2057L} {697, 2057}

\bibitem[\protect\citeauthoryear{{Li}, {Gao}, {Ding}, {Wang}  \& {Zhang}}{{Li}
  et~al.}{2018}]{Li2018}
{Li} Z.-X.,  {Gao} H.,  {Ding} X.-H.,  {Wang} G.-J.,   {Zhang} B.,  2018,
  \mn@doi [Nature Communications] {10.1038/s41467-018-06303-0}, \href
  {https://ui.adsabs.harvard.edu/abs/2018NatCo...9.3833L} {9, 3833}

\bibitem[\protect\citeauthoryear{{Lister}}{{Lister}}{2003}]{Lister2003}
{Lister} M.~L.,  2003, \mn@doi [\apj] {10.1086/379241}, \href
  {https://ui.adsabs.harvard.edu/abs/2003ApJ...599..105L} {599, 105}

\bibitem[\protect\citeauthoryear{{Lister} \& {Homan}}{{Lister} \&
  {Homan}}{2005}]{Lister2005}
{Lister} M.~L.,  {Homan} D.~C.,  2005, \mn@doi [\aj] {10.1086/432969}, \href
  {https://ui.adsabs.harvard.edu/abs/2005AJ....130.1389L} {130, 1389}

\bibitem[\protect\citeauthoryear{{Lister} \& {Marscher}}{{Lister} \&
  {Marscher}}{1997}]{Lister1997}
{Lister} M.~L.,  {Marscher} A.~P.,  1997, \mn@doi [\apj] {10.1086/303629},
  \href {https://ui.adsabs.harvard.edu/abs/1997ApJ...476..572L} {476, 572}

\bibitem[\protect\citeauthoryear{{Lister} et~al.,}{{Lister}
  et~al.}{2016}]{Lister2016}
{Lister} M.~L.,  et~al., 2016, \mn@doi [\aj] {10.3847/0004-6256/152/1/12},
  \href {https://ui.adsabs.harvard.edu/abs/2016AJ....152...12L} {152, 12}

\bibitem[\protect\citeauthoryear{{Lister} et~al.,}{{Lister}
  et~al.}{2019}]{Lister2019}
{Lister} M.~L.,  et~al., 2019, \mn@doi [\apj] {10.3847/1538-4357/ab08ee}, \href
  {https://ui.adsabs.harvard.edu/abs/2019ApJ...874...43L} {874, 43}

\bibitem[\protect\citeauthoryear{{Lister}, {Homan}, {Kellermann}, {Kovalev},
  {Pushkarev}, {Ros}  \& {Savolainen}}{{Lister} et~al.}{2021}]{Lister2021}
{Lister} M.~L.,  {Homan} D.~C.,  {Kellermann} K.~I.,  {Kovalev} Y.~Y.,
  {Pushkarev} A.~B.,  {Ros} E.,   {Savolainen} T.,  2021, \mn@doi [\apj]
  {10.3847/1538-4357/ac230f}, \href
  {https://ui.adsabs.harvard.edu/abs/2021ApJ...923...30L} {923, 30}

\bibitem[\protect\citeauthoryear{{Lu} \& {Qin}}{{Lu} \& {Qin}}{2021}]{Lu2021}
{Lu} W.-J.,  {Qin} Y.-P.,  2021, arXiv e-prints, \href
  {https://ui.adsabs.harvard.edu/abs/2021arXiv210705575L} {p. arXiv:2107.05575}

\bibitem[\protect\citeauthoryear{{Lynden-Bell}}{{Lynden-Bell}}{1977}]{Lynden1977}
{Lynden-Bell} D.,  1977, \mn@doi [\nat] {10.1038/270396a0}, \href
  {https://ui.adsabs.harvard.edu/abs/1977Natur.270..396L} {270, 396}

\bibitem[\protect\citeauthoryear{Massey}{Massey}{1951}]{KS}
Massey F.~J.,  1951, Journal of the American Statistical Association, 46, 68

\bibitem[\protect\citeauthoryear{{McKinney} \& {Gammie}}{{McKinney} \&
  {Gammie}}{2004}]{McKinney2004}
{McKinney} J.~C.,  {Gammie} C.~F.,  2004, \mn@doi [\apj] {10.1086/422244},
  \href {https://ui.adsabs.harvard.edu/abs/2004ApJ...611..977M} {611, 977}

\bibitem[\protect\citeauthoryear{{Mirabel} \& {Rodr{\'\i}guez}}{{Mirabel} \&
  {Rodr{\'\i}guez}}{1994}]{Mirabel1994}
{Mirabel} I.~F.,  {Rodr{\'\i}guez} L.~F.,  1994, \mn@doi [\nat]
  {10.1038/371046a0}, \href
  {https://ui.adsabs.harvard.edu/abs/1994Natur.371...46M} {371, 46}

\bibitem[\protect\citeauthoryear{{Planck Collaboration} et~al.,}{{Planck
  Collaboration} et~al.}{2020}]{Planck2020}
{Planck Collaboration} et~al., 2020, \mn@doi [\aap]
  {10.1051/0004-6361/201833910}, \href
  {https://ui.adsabs.harvard.edu/abs/2020A&A...641A...6P} {641, A6}

\bibitem[\protect\citeauthoryear{{Poulin}, {Smith}, {Karwal}  \&
  {Kamionkowski}}{{Poulin} et~al.}{2019}]{Poulin2019}
{Poulin} V.,  {Smith} T.~L.,  {Karwal} T.,   {Kamionkowski} M.,  2019, \mn@doi
  [\prl] {10.1103/PhysRevLett.122.221301}, \href
  {https://ui.adsabs.harvard.edu/abs/2019PhRvL.122v1301P} {122, 221301}

\bibitem[\protect\citeauthoryear{{Pracy} et~al.,}{{Pracy}
  et~al.}{2016}]{Pracy2016}
{Pracy} M.~B.,  et~al., 2016, \mn@doi [\mnras] {10.1093/mnras/stw910}, \href
  {https://ui.adsabs.harvard.edu/abs/2016MNRAS.460....2P} {460, 2}

\bibitem[\protect\citeauthoryear{{Qin}}{{Qin}}{1999}]{Qin1999}
{Qin} Y.-P.,  1999, \mn@doi [Modern Physics Letters A]
  {10.1142/S0217732399001140}, \href
  {https://ui.adsabs.harvard.edu/abs/1999MPLA...14.1073Q} {14, 1073}

\bibitem[\protect\citeauthoryear{{Rees}}{{Rees}}{1966}]{Rees1966}
{Rees} M.~J.,  1966, \mn@doi [\nat] {10.1038/211468a0}, \href
  {https://ui.adsabs.harvard.edu/abs/1966Natur.211..468R} {211, 468}

\bibitem[\protect\citeauthoryear{{Refsdal}}{{Refsdal}}{1964}]{Refsdal1964}
{Refsdal} S.,  1964, \mn@doi [\mnras] {10.1093/mnras/128.4.307}, \href
  {https://ui.adsabs.harvard.edu/abs/1964MNRAS.128..307R} {128, 307}

\bibitem[\protect\citeauthoryear{{Riess} et~al.,}{{Riess}
  et~al.}{2021}]{Riess2021}
{Riess} A.~G.,  et~al., 2021, arXiv e-prints, \href
  {https://ui.adsabs.harvard.edu/abs/2021arXiv211204510R} {p. arXiv:2112.04510}

\bibitem[\protect\citeauthoryear{{Schutz}}{{Schutz}}{1986}]{Schutz1986}
{Schutz} B.~F.,  1986, \mn@doi [\nat] {10.1038/323310a0}, \href
  {https://ui.adsabs.harvard.edu/abs/1986Natur.323..310S} {323, 310}

\bibitem[\protect\citeauthoryear{{Seto} \& {Toda}}{{Seto} \&
  {Toda}}{2021}]{Seto2021}
{Seto} O.,  {Toda} Y.,  2021, \mn@doi [\prd] {10.1103/PhysRevD.103.123501},
  \href {https://ui.adsabs.harvard.edu/abs/2021PhRvD.103l3501S} {103, 123501}

\bibitem[\protect\citeauthoryear{{Sparks}, {Fraix-Burnet}, {Macchetto}  \&
  {Owen}}{{Sparks} et~al.}{1992}]{Sparks1992}
{Sparks} W.~B.,  {Fraix-Burnet} D.,  {Macchetto} F.,   {Owen} F.~N.,  1992,
  \mn@doi [\nat] {10.1038/355804a0}, \href
  {https://ui.adsabs.harvard.edu/abs/1992Natur.355..804S} {355, 804}

\bibitem[\protect\citeauthoryear{{Taylor} \& {Vermeulen}}{{Taylor} \&
  {Vermeulen}}{1997}]{Taylor1997}
{Taylor} G.~B.,  {Vermeulen} R.~C.,  1997, \mn@doi [\apjl] {10.1086/310800},
  \href {https://ui.adsabs.harvard.edu/abs/1997ApJ...485L...9T} {485, L9}

\bibitem[\protect\citeauthoryear{{Vega-Ferrero}, {Diego}, {Miranda}  \&
  {Bernstein}}{{Vega-Ferrero} et~al.}{2018}]{Ferrero2018}
{Vega-Ferrero} J.,  {Diego} J.~M.,  {Miranda} V.,   {Bernstein} G.~M.,  2018,
  \mn@doi [\apjl] {10.3847/2041-8213/aaa95f}, \href
  {https://ui.adsabs.harvard.edu/abs/2018ApJ...853L..31V} {853, L31}

\bibitem[\protect\citeauthoryear{{Verde}, {Treu}  \& {Riess}}{{Verde}
  et~al.}{2019}]{Verde2019}
{Verde} L.,  {Treu} T.,   {Riess} A.~G.,  2019, \mn@doi [Nature Astronomy]
  {10.1038/s41550-019-0902-0}, \href
  {https://ui.adsabs.harvard.edu/abs/2019NatAs...3..891V} {3, 891}

\bibitem[\protect\citeauthoryear{{Weaver} et~al.,}{{Weaver}
  et~al.}{2022}]{Weaver2022}
{Weaver} Z.~R.,  et~al., 2022, \mn@doi [\apjs] {10.3847/1538-4365/ac589c},
  \href {https://ui.adsabs.harvard.edu/abs/2022ApJS..260...12W} {260, 12}

\bibitem[\protect\citeauthoryear{{Wilkinson}, {Readhead}, {Purcell}  \&
  {Anderson}}{{Wilkinson} et~al.}{1977}]{Wilkinson1977}
{Wilkinson} P.~N.,  {Readhead} A.~C.~S.,  {Purcell} G.~H.,   {Anderson} B.,
  1977, \mn@doi [\nat] {10.1038/269764a0}, \href
  {https://ui.adsabs.harvard.edu/abs/1977Natur.269..764W} {269, 764}

\bibitem[\protect\citeauthoryear{{Wong} et~al.,}{{Wong}
  et~al.}{2020}]{Wong2020}
{Wong} K.~C.,  et~al., 2020, \mn@doi [\mnras] {10.1093/mnras/stz3094}, \href
  {https://ui.adsabs.harvard.edu/abs/2020MNRAS.498.1420W} {498, 1420}

\bibitem[\protect\citeauthoryear{{Wu}, {Zhang}  \& {Wang}}{{Wu}
  et~al.}{2021}]{Wu2021}
{Wu} Q.,  {Zhang} G.~Q.,   {Wang} F.~Y.,  2021, arXiv e-prints, \href
  {https://ui.adsabs.harvard.edu/abs/2021arXiv210800581W} {p. arXiv:2108.00581}

\bibitem[\protect\citeauthoryear{{Yin}}{{Yin}}{2020}]{Yin2020}
{Yin} L.,  2020, arXiv e-prints, \href
  {https://ui.adsabs.harvard.edu/abs/2020arXiv201213917Y} {p. arXiv:2012.13917}

\bibitem[\protect\citeauthoryear{{Yuan}, {Wang}, {Worrall}, {Zhang}  \&
  {Mao}}{{Yuan} et~al.}{2018}]{Yuan2018}
{Yuan} Z.,  {Wang} J.,  {Worrall} D.~M.,  {Zhang} B.-B.,   {Mao} J.,  2018,
  \mn@doi [\apjs] {10.3847/1538-4365/aaed3b}, \href
  {https://ui.adsabs.harvard.edu/abs/2018ApJS..239...33Y} {239, 33}

\bibitem[\protect\citeauthoryear{{van Velzen}, {Falcke}, {Schellart},
  {Nierstenh{\"o}fer}  \& {Kampert}}{{van Velzen} et~al.}{2012}]{Velzen2012}
{van Velzen} S.,  {Falcke} H.,  {Schellart} P.,  {Nierstenh{\"o}fer} N.,
  {Kampert} K.-H.,  2012, \mn@doi [\aap] {10.1051/0004-6361/201219389}, \href
  {https://ui.adsabs.harvard.edu/abs/2012A&A...544A..18V} {544, A18}

\makeatother
\end{thebibliography}
\bsp	
\label{lastpage}
\end{document}